\documentclass{article}
\usepackage[utf8]{inputenc}
\usepackage{tikz}
\usepackage{blkarray}

 \usepackage[normalem]{ulem}
 \usepackage{cite}

\usepackage[english]{babel} 
\usepackage{graphicx}
\usepackage{float}

\usepackage{authblk}

\newmuskip\pFqskip
\mathchardef\pFcomma=\mathcode`, 

\newcommand*\pFq[5]{%
  \begingroup
  \begingroup\lccode`~=`,
    \lowercase{\endgroup\def~}{\pFcomma\mkern\pFqskip}%
  \mathcode`,=\string"8000
  {}_{#1}F_{#2}\biggl[\genfrac..{0pt}{}{#3}{#4};#5\biggr]%
  \endgroup
}

\usepackage{mathtools}

\DeclarePairedDelimiterX\phys[3]{\langle}{\rangle}{#1 \delimsize\vert\mathopen{} #2 \delimsize\vert\mathopen{} #3}

\usepackage{caption}
\usepackage{subcaption}

\usepackage{bm,amsmath,amssymb,comment,esint,accents,physics}
\usepackage{dsfont,enumerate}

\usepackage{tikz}

\usepackage{soul}
\usepackage{color}

\usepackage[margin=0.75in]{geometry}

\newcommand{\tx}[1]{\textrm{#1}}

\newcommand{\lrp}[1]{\left(#1\right)}

\makeatletter
\newsavebox{\@brx}
\newcommand{\llangle}[1][]{\savebox{\@brx}{\(\m@th{#1\langle}\)}%
  \mathopen{\copy\@brx\kern-0.5\wd\@brx\usebox{\@brx}}}
\newcommand{\rrangle}[1][]{\savebox{\@brx}{\(\m@th{#1\rangle}\)}%
  \mathclose{\copy\@brx\kern-0.5\wd\@brx\usebox{\@brx}}}
\makeatother

\newcommand{\A}{\mathcal{A}}

\usepackage[colorlinks,bookmarks=false,citecolor=red,linkcolor=blue,hyperfootnotes=true,urlcolor=magenta]{hyperref}
\begin{document}

\title{\bf Absence of logarithmic enhancement in {the} entanglement scaling {of} free fermions on folded {cubes}}

\author[1]{Pierre-Antoine Bernard}

\author[2]{Zachary Mann}

\author[1]{Gilles Parez}

\author[1,3]{Luc Vinet}

\affil[1]{\it Centre de Recherches Math\'ematiques (CRM), Universit\'e de Montr\'eal, P.O. Box 6128, Centre-ville
Station, Montr\'eal (Qu\'ebec), H3C 3J7, Canada}
\affil[2]{\it Institute for Quantum Computing, University of Waterloo, Waterloo N2L 3G1, Ontario, Canada \color{black}}
\affil[3]{\it IVADO, 6666 Rue Saint-Urbain, Montr\'eal (Qu\'ebec), H2S 3H1, Canada}

\date{\today}

\maketitle
\begin{abstract}
    This study investigates the scaling behavior of the ground-state entanglement entropy in a model of free fermions on folded cubes. An analytical expression is derived in the large-diameter limit, revealing a strict adherence to the area law. The absence of the logarithmic enhancement expected for free fermions is explained using a decomposition of folded cubes in chains based on its Terwilliger algebra and $\mathfrak{so}(3)_{-1}$. The entanglement Hamiltonian and its relation to Heun operators are also investigated.
\end{abstract}

\section{Introduction}

Key physical attributes of quantum many-body systems can be studied from their thermodynamic properties in the limit of infinite number of degrees of freedom. For instance, understanding the scaling behavior of the entanglement entropy is crucial \cite{amico2008entanglement,laflorencie2016quantum}, as it provides a mean to detect and characterize quantum phase transitions \cite{OAFF02,ON02,vidal2003entanglement,CC04,calabrese2009entanglement}, probe topological phases of matter \cite{kitaev2006topological,levin2006detecting} and investigate the emergence of thermodynamics in non-equilibrium situations~\cite{cc-05,fc-08,GE15,ac-17}.

For gapped bipartite systems $A \cup B$, the ground-state entanglement entropy $S_A$ of a region $A$ typically obeys an area law \cite{amico2008entanglement, CC04}. That is, the entanglement entropy scales with the size of the boundary $\partial A$ between $A$ and its complement $B$,
\begin{equation}
    S_A \sim |\partial A |.
\end{equation}

In contrast, for critical 
free fermions on cubic lattices in arbitrary dimensions, the scaling of the entanglement entropy exhibits a logarithmic violation, or enhancement, of the area law \cite{gioev2006entanglement,li2006scaling},
\begin{equation}
    S_A \sim |\partial A | \ln{\ell},
\end{equation}
where $\ell$ is the size of region $A$. Most notably, for one-dimensional quantum critical models described by a $1+1d$ conformal field theory (CFT) in the scaling limit, the entanglement entropy of an interval of length $\ell$ embedded in an infinite chain reads \cite{CC04}
\begin{equation}
    S_A = \frac c3 \ln \ell + \dots \ ,
\end{equation}
where $c$ is the central charge of the underlying CFT.

The situation becomes {more intricate} 
as we 
consider thermodynamic limits based on more general sequences of graphs with an increasing number of sites. The scaling of the entanglement entropy and multipartite information was considered recently for free fermions hopping on the vertices of Hamming \cite{bernard2023entanglement, parez2022multipartite} and Johnson graphs \cite{bernard2021entanglement} in the large-diameter
limit. In these cases, it was observed that $S_A$ {either respects the area law, or exhibits a logarithmic suppression thereof. The lack of logarithmic enhancement for free-fermion models defined on these graphs}
suggests that there is a strong interplay between the {geometry of the underlying lattice}
of a many-body system and the entanglement content of its ground state. 

Motivated by these observations, this work focuses on a model of free fermions hopping on the vertices of a folded cube. Specifically, an analytical expression is derived for the entanglement entropy in the large-diameter limit, revealing the absence of a logarithmic enhancement {of the area law}. An explanation is given in terms of the diameter of the graph perceived by the degrees of freedom on the boundary $\partial A$. 

The structure of the paper is as follows. In Sec. \ref{s2}, the model of free fermions on folded cubes is introduced. In Sec. \ref{s3}, we recall the concept of entanglement entropy and an analytical formula is obtained for the model of interest. The derivation is based on the relation between the Terwilliger algebra of folded cubes and the algebra $\mathfrak{so}(3)_{-1}$ \cite{brown2013hypercubes,t1,t2,t3} which enables an effective decomposition of the system into a direct sum of independent free-fermion systems on inhomogeneous chains. The absence of logarithmic enhancement is discussed from the perspective of this decomposition. Section \ref{s4} investigates the relation between the entanglement Hamiltonian and algebraic Heun operators based on $\mathfrak{so}(3)_{-1}$ generators. This type of relation was first considered for fermions on homogeneous lattices in \cite{eisler2013free,eisler2017analytical,eisler2018properties}. In the case of folded cubes, we demonstrate through numerical computations that approximating the entanglement Hamiltonian with an affine transformation of the Heun operator reproduces the R\'enyi entropies of the reduced density matrix with great precision. R\'eyni fidelities \cite{parez2022symmetry} between the two respective density matrices are also computed numerically to quantify the accuracy of the approximation. We offer our concluding remarks and outlooks in Sec.~\ref{sec:conclusion}.

\section{Free fermions on folded cubes}
\label{s2}
\subsection{$d$-cubes and folded $d$-cubes }
\begin{figure}
\begin{subfigure}{.5\textwidth}
\centering
    \begin{tikzpicture}
\draw (-2.5,0) -- (-1,1.5);
\draw (-2.5,0) -- (-0.5,0);
\draw (-2.5,0) -- (-1,-1.5);

\draw (2.5,0) -- (1,1.5);
\draw (2.5,0) -- (0.5,0);
\draw (2.5,0) -- (1,-1.5);

\draw (-1,1.5) -- (1,1.5);
\draw (-1,1.5) -- (0.5,0);

\draw (-1,-1.5) -- (0.5,0);
\draw (-1,-1.5) -- (1,-1.5);

\draw (-0.5,0) -- (1,1.5);
\draw (-0.5,0) -- (1,-1.5);

\draw[black,fill=white] (1,1.5) circle (0.13cm);
\node[] at (1.7,1.7) {(1,1,0)};
\draw[black,fill=white] (0.5,0) circle (0.13cm);
\node[] at (1.3,0.3) {(0,1,1)};
\draw[black,fill=white] (1,-1.5) circle (0.13cm);
\node[] at (1.7,-1.5) {(1,0,1)};

\draw[black,fill=white] (-1,1.5) circle (0.13cm);
\node[] at (-1.7,1.7) {(0,1,0)};
\draw[black,fill=white] (-0.5,0) circle (0.13cm);
\node[] at (-1.3,0.3) {(1,0,0)};
\draw[black,fill=white] (-1,-1.5) circle (0.13cm);
\node[] at (-1.7,-1.5) {(0,0,1)};

\draw[black,fill=white] (2.5,0) circle (0.13cm);
\node[] at (3.2,0.2) {(1,1,1)};
\draw[black,fill=white] (-2.5,0) circle (0.13cm);
\node[] at (-3.2,0.2) {(0,0,0)};
\end{tikzpicture}
    \label{fig:subham}
\end{subfigure}%
\begin{subfigure}{.5\textwidth}
 \centering
    \begin{tikzpicture}[scale = 0.8]
\draw (-2.5,0) -- (-2.5-5,0-1);
\draw (2.5,0) -- (2.5-5,0-1);
\draw (-1,-1.5) -- (-1-5,-1.5-1);
\draw (-0.5,0) -- (-0.5-5,0-1);
\draw (1,1.5) -- (1-5,1.5-1);
\draw (-0.5,0)  -- (-0.5-5,0-1);
\draw (1,-1.5) -- (1-5,-1.5-1);
\draw (-1,1.5) -- (-1-5,1.5-1);

\draw (-2.5-5,0-1) -- (-1-5,1.5-1);
\draw (-2.5-5,0-1) -- (-0.5-5,0-1);
\draw (-2.5-5,0-1) -- (-1-5,-1.5-1);

\draw (2.5-5,0-1) -- (1-5,1.5-1);
\draw (2.5-5,0-1) -- (0.5-5,0-1);
\draw (2.5-5,0-1) -- (1-5,-1.5-1);

\draw (-1-5,1.5-1) -- (1-5,1.5-1);
\draw (-1-5,1.5-1) -- (0.5-5,0-1);

\draw (-1-5,-1.5-1) -- (0.5-5,0-1);
\draw (-1-5,-1.5-1) -- (1-5,-1.5-1);

\draw (-0.5-5,0-1) -- (1-5,1.5-1);
\draw (-0.5-5,0-1) -- (1-5,-1.5-1);

\draw[black,fill=white] (1-5,1.5-1) circle (0.13cm);
\draw[black,fill=white] (0.5-5,0-1) circle (0.13cm);
\draw[black,fill=white] (1-5,-1.5-1) circle (0.13cm);

\draw[black,fill=white] (-1-5,1.5-1) circle (0.13cm);
\draw[black,fill=white] (-0.5-5,0-1) circle (0.13cm);
\draw[black,fill=white] (-1-5,-1.5-1) circle (0.13cm);

\draw[black,fill=white] (2.5-5,0-1) circle (0.13cm);
\draw[black,fill=white] (-2.5-5,0-1) circle (0.13cm);


\draw[line width=0.25mm, black ] (-2.5,0) -- (-1,1.5);
\draw (-2.5,0) -- (-0.5,0);
\draw[line width=0.25mm, black ] (-2.5,0) -- (-1,-1.5);

\draw (2.5,0) -- (1,1.5);
\draw (2.5,0) -- (0.5,0);
\draw (2.5,0) -- (1,-1.5);

\draw (-1,1.5) -- (1,1.5);
\draw[line width=0.25mm, black ] (-1,1.5) -- (0.5,0);

\draw[line width=0.25mm, black ] (-1,-1.5) -- (0.5,0);
\draw (-1,-1.5) -- (1,-1.5);

\draw (-0.5,0) -- (1,1.5);
\draw (-0.5,0) -- (1,-1.5);

\draw[black,fill=white] (1,1.5) circle (0.13cm);
\draw[black,fill=white] (0.5,0) circle (0.13cm);
\draw[black,fill=white] (1,-1.5) circle (0.13cm);

\draw[black,fill=white] (-1,1.5) circle (0.13cm);
\draw[black,fill=white] (-0.5,0) circle (0.13cm);
\draw[black,fill=white] (-1,-1.5) circle (0.13cm);

\draw[black,fill=white] (2.5,0) circle (0.13cm);
\draw[black,fill=white] (-2.5,0) circle (0.13cm);

\end{tikzpicture}
    \label{fig:subham}
\end{subfigure}
\caption{A $3$-cube (left) and a $4$-cube (right).}
\label{fig:cubes}
\end{figure}
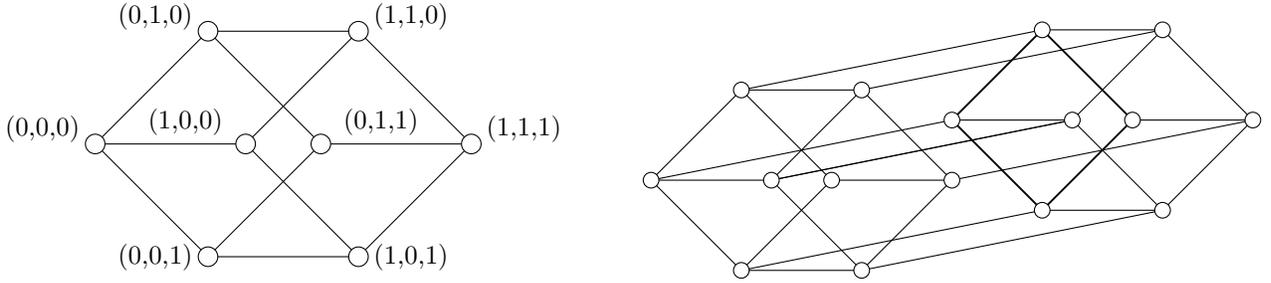

The $d$-cube or hypercube graph $H(d,2)$ has a 
set of vertices $X_d$ given by the binary strings of length $d$,
\begin{equation}
    X_d = \{ v = (v_1, v_2, \dots, v_d) \ | \  v_i \in \{0,1\}\}.
\end{equation}
An edge connects two vertices $v$ and $v'$ if there is a unique position $i$ such that they differ, i.e. $v_i \neq v_i'$. We illustrate a $3$- and $4$-cube in Fig. \ref{fig:cubes}. For an arbitrary pair of vertices $v$ and $v'$, their relative distance is given by the Hamming distance,
\begin{equation}
    \partial(v,v') = |\{ i \in \{1,2,\dots, d\}\ | \ v_i \neq v_i'\}|.
\end{equation}

The ground-state entanglement properties of free fermions defined on such lattices have been investigated, and results regarding bipartite entanglement and multipartite information have been obtained in \cite{bernard2023entanglement,parez2022multipartite}.

Folded $d$-cubes, denoted $\square_d$, are obtained by taking the antipodal quotient of a $d$-cube, i.e. by merging the pairs of vertices at distance $d$ in $H(d,2)$ \cite{brouwer2012distance}. The vertices of  $\square_d$ are the equivalence classes in $ X_d/\sim $, where the relation~$\sim$ is defined by
\begin{equation}
    v \sim v' \iff \partial(v,v') \in \{0,d\}.
\end{equation}

\begin{figure}
\centering
    \begin{tikzpicture}
\draw (-2,-2) -- (-2,2);
\draw (-2,2) -- (2,2);
\draw (2,2) -- (2,-2);
\draw (2,-2) -- (-2,-2);

\draw (-1,-1) -- (-1,1);
\draw (-1,1) -- (1,1);
\draw (1,1) -- (1,-1);
\draw (1,-1) -- (-1,-1);

\draw[color = red] (-2,-2) -- (-1,1);
\draw (-2,-2) -- (1,-1);

\draw (2,-2) -- (1,1);
\draw[color = red] (2,-2) -- (-1,-1);

\draw (-2,2) -- (-1,-1);
\draw[color = red] (-2,2) -- (1,1);

\draw[color = red] (2,2) -- (1,-1);
\draw (2,2) -- (-1,1);

\draw[black,fill=white] (2,2) circle (0.13cm);
\draw[black,fill=white] (-2,-2) circle (0.13cm);
\draw[black,fill=white] (2,-2) circle (0.13cm);
\draw[black,fill=white] (-2,2) circle (0.13cm);

\draw[black,fill=white] (1,1) circle (0.13cm);
\draw[black,fill=white] (-1,-1) circle (0.13cm);
\draw[black,fill=white] (1,-1) circle (0.13cm);
\draw[black,fill=white] (-1,1) circle (0.13cm);
\end{tikzpicture}
    \label{fig:subham}
    \caption{Folded $4$-cube. It is either obtained by merging antipodal vertices in a $4$-cube or by adding the red edges between the antipodal vertices of the $3$-cube in black.}
\label{fig:Fcubes}
\end{figure}
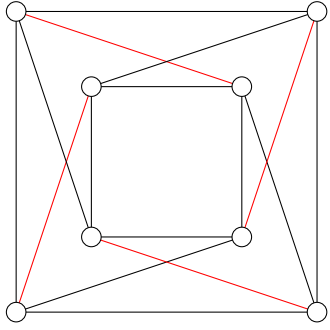

Two classes $[v]$ and $[v']$ are connected by an edge in $\square_d$ if their respective representative strings $v$ and $v'$ are at a Hamming distance $1$ or $d-1$ in $H(d,2)$. The graph $\square_d$ is also obtained by taking the vertices $  X_{d-1} = X_d/\sim$ of a $(d-1)$-cube and connecting by an edge those at Hamming distance $1$ or $d-1$. The set $E$ of edges of $\square_d$ is thus
\begin{equation}
    E = \{ (v,v') \in X_{d-1} \times X_{d-1} \ | \ \partial(v,v') \in \{1,d-1\}\},
\end{equation}
and the distance between any two vertices in $\square_d$ is given by
\begin{equation}
    \text{dist}(v,v') = \text{min}\{ \partial(v,v'), d - \partial(v,v')\}.
    \label{distfold}
\end{equation}
{We illustrate a folded $4$-cube in Fig.~\ref{fig:Fcubes}.}

Each vertex $v = (v_1, \dots, v_{d-1}) \in X_{d-1}$ can be associated with a vector in $\mathbb{C}^{2^{d-1}}$ 
 {in} the following way,
\begin{equation}
    \ket{v} = \ket{v_1}\otimes\ket{v_2}\otimes \dots \otimes \ket{v_{d-1}},
    \label{bas1}
\end{equation}
where $\ket{0} = (1,0)^t$ and $\ket{1} = (0,1)^t$. In this basis, the adjacency matrix $\A$ of $\square_d$ is given by
\begin{equation}
    \A = \left(\sum_{n = 1}^{d-1} \underbrace{I \otimes I \otimes ... \otimes I}_{n-1 \text{ times}} \otimes \ \sigma_x \otimes \underbrace{I \otimes ... \otimes I}_{d-1-n \text{ times}}\right) + \underbrace{\ \sigma_x \otimes \ \sigma_x \otimes ... \otimes \ \sigma_x}_{d-1 \text{ times}},
    \label{adj}
\end{equation}
where $I$ is the $2 \times 2$ identity matrix and $\sigma_x$ is a Pauli matrix. Indeed, one can check that
\begin{equation}
    \bra{v}\A \ket{v'} = \left\{
	\begin{array}{ll}
		1  & \mbox{if } (v,v') \in E \\
		0 & \mbox{otherwise. }
	\end{array}
\right.
\end{equation}
 
Let us note that the distance function \eqref{distfold} implies that both folded cubes $\square_{2d}$ and $\square_{2d +1}$ have diameter $d$. In the following, we restrict ourselves to the case $\square_{2d +1}$.

\subsection{Free-fermion Hamiltonian on $\square_{2d + 1}$}

For each vertex $v \in X_{2d}$, we define a pair of fermionic creation and annihilation operators $c_v^\dagger$ and $c_v$ that verify the canonical anti-commutation relations
\begin{equation}
     \{c_v, c_{v'}\} = \{c_v^\dagger, c_{v'}^\dagger\} = 0, \quad \{c_v, c_{v'}^\dagger\} = \delta_{vv'}.
\end{equation}

A nearest-neighbor Hamiltonian on the graph $\square_{2d+1}$ is then defined as
\begin{equation}\label{eq:H}
\begin{split}
        \mathcal{H} &:= \sum_{(v,v') \in E} c_v^\dagger c_{v'} = \boldsymbol{c}^\dagger \A \boldsymbol{c},
\end{split}
\end{equation}
where $\A$ is the adjacency matrix of $\square_{2d+1}$, and
\begin{equation}
     \boldsymbol{c} = \sum_{v \in X_{2d}} c_v \ket{v}, \quad \boldsymbol{c}^\dagger = \sum_{v \in X_{2d}} c_v^\dagger \bra{v}.
\end{equation}\\

The Hamiltonian $\mathcal{H}$ can be diagonalized using the eigenvectors of 
the adjacency matrix $\A$. These are given by the $2d$-fold tensor product of the eigenvectors $\ket{\pm}$ of $\sigma_x$,
\begin{equation}
    \ket{\pm} =\frac{1}{\sqrt{2}}
    \begin{pmatrix}
1 \\
\pm1  
\end{pmatrix}, \quad \sigma_x \ket{\pm} = \pm \ket{\pm}.
\end{equation}
{The eigenvectors of $\A$ are denoted} 
$\ket{\theta_k, \ell}$, where $k$ indicates the number of vector $\ket{+}$ in the tensor product and $\ell \in \{1, 2, \dots, \binom{2d}{k}\}$ is a label for the degeneracy,
\begin{equation}
    \A \ket{\theta_k, \ell} = \theta_k \ket{\theta_k, \ell}, \quad \theta_k =  2k - 2d + (-1)^{k}.
\end{equation}

The degeneracy of the model is not entirely captured by the index $\ell$ since we also have $\theta_{2k} = \theta_{2k + 1}$ for all $k \in \{0, \dots, d-1\}$. In terms of $\theta_k$ and the eigenvectors $\ket{\theta_k, \ell}$, the free-fermion Hamiltonian $\mathcal{H}$ can be rewritten as
\begin{equation}
    \mathcal{H} = \sum_{k = 0}^{2d} \sum_{\ell = 1}^{\binom{2d}{k}} \theta_k \hat{c}_{k \ell}^\dagger \hat{c}_{k \ell},
\end{equation}
where $\hat{c}_{k \ell}$ and $\hat{c}_{k \ell}^\dagger$ are fermionic creation and annihilation operators which verify the same canonical anti-commutation relations as $c_v$, $c_v^\dagger$, and are defined by
\begin{equation}
    \hat{c}_{k \ell} = \sum_{v \in X_{2d}} \bra{\theta_k,\ell}\ket{v}c_v , \quad  \hat{c}_{k \ell}^\dagger = \sum_{v \in X_{2d}}  \bra{v}\ket{\theta_k, \ell}c_v^\dagger.
\end{equation}

\subsection{Ground state and correlation matrix}
Let $| 0 \rrangle$ be the vacuum state which {is annihilated by}
all operators $c_v$, 
\begin{equation}
    {c}_{v}| 0 \rrangle=0,\quad\forall\ v \in X_{2d}.
\end{equation}

The ground state $|\Psi_0\rrangle$ of $\mathcal{H}$ is obtained by acting on the vacuum with all creation operators $\hat{c}^\dagger_{k \ell}$ associated with negative energy excitation, i.e. $\theta_k < 0$. By filling up the Fermi sea, we have
\begin{equation}
\label{eq:psi0}
    |\Psi_0\rrangle =\lrp{\prod_{k = 0}^{2K+1}\prod_{\ell=1}^{\binom{2d}{k}} \hat{c}^{\dagger}_{k \ell}}|0\rrangle,
\end{equation}
where $K$ is the integer such that $\theta_{2K} =\theta_{2K+1}  < 0$ and $\theta_{2K+2} \geqslant 0$. A direct computation shows that {the two-point correlation matrix in this state is}
\begin{equation}
  \hat{C}_{vv'} = \llangle \Psi_0 | c_v^\dagger c_{v'}  |\Psi_0\rrangle = \sum_{k = 0}^{2K + 1} \sum_{\ell = 1}^{\binom{2d}{k}} \bra{v}\ket{\theta_k, \ell} \bra{\theta_k, \ell}\ket{v'}.
\end{equation}

For any distance $n := \text{dist}(v,v')$, a combinatorial argument allows to {express} the correlation function as
\begin{equation}
  \hat{C}_{vv'} = \frac{1}{2^{2d}} \sum_{k = 0}^{N}\binom{2d-n}{k} \pFq{2}{1}{-n, -k}{2d-n-k+1}{-1} + \frac{1}{2^{2d}} \sum_{k = N + 1}^{ 2K + 1}\binom{n}{2d -k} \pFq{2}{1}{n - 2d, \ k-2d}{n+k - 2d +1}{-1},
\end{equation}
where $N = \text{min}(2d-n, 2K + 1)$ and ${}_2F_1$ is Gauss hypergeometric function. Let us also note that the correlation matrix $\hat{C}$ whose entries are $\hat{C}_{vv'}$ can be expressed as a sum of projectors $E_{2k} + E_{2k+1}$ onto eigenspaces of $\A$,
\begin{equation}
    \hat{C} = \sum_{k =0}^K \left(E_{2k} + E_{2k +1}\right), \quad E_k = \sum_{\ell = 1}^{\binom{2d}{k}}\ket{\theta_k, \ell} \bra{\theta_k, \ell}.
\end{equation}

\section{Bipartite entanglement entropy}
\label{s3}

{In this section we investigate the ground-state entanglement entropy of free fermions defined on the folded cube.} 

\subsection{Definitions}

{For a given bipartition $A \cup B$ of a quantum many-body system in a pure state $|\Psi_0\rrangle$, the entanglement entropy is defined as the}
von Neumann entropy of the reduced density matrix $\rho_A$,
\begin{equation}
 S_A = - \text{tr}_A\left(\rho_A \ln{\rho_A}\right), \quad \rho_A =   \text{tr}_B (|\Psi_0\rrangle \llangle \Psi_0|).
\end{equation}

We are interested in computing this quantity for free fermions on folded cubes $\square_{2d +1}$, in their ground state $|\Psi_0\rrangle$ defined in \eqref{eq:psi0}. Since this state is a Slater determinant, {the reduced density matrix $\rho_A$ is a Gaussian operator,}
\begin{equation}
    \rho_A = \frac{1}{\mathcal{K}}e^{- \mathcal{H}_{\text{ent}}}, \quad \mathcal{K} = \text{tr}_A\left(e^{- \mathcal{H}_{\text{ent}}}\right),
\end{equation}
where the entanglement Hamiltonian $\mathcal{H}_{\text{ent}}$ is quadratic in fermionic operators associated to degrees of freedom in region $A$,
\begin{equation}
    \mathcal{H}_{\text{ent}} = \sum_{v,v'  \in A} h_{vv'} c_v^\dagger c_{v'}.
    \label{enthamil}
\end{equation}

Owning to the quadratic nature of the free-fermion Hamiltonian, one can relate the matrix $h$ in the definition of the entanglement Hamiltonian \eqref{enthamil} to the the restriction of the correlation matrix $\hat{C}$ to region~$A$~\cite{peschel2003calculation},
\begin{equation}
    h =  \ln\left(\frac{1-C}{C}\right),
\end{equation}
where $C$ is referred to as the truncated correlation matrix. It is defined as
\begin{equation}
    C = \pi_A \hat{C} \pi_A, \quad \pi_A = \sum_{v\in A } \ket{v}\bra{v},
\end{equation}
with $\pi_A$ the projector in $\mathbb{C}^{2d}$ onto the vector space associated to sites in the region $A$. It follows that the entanglement entropy is given in terms of the eigenvalues $\lambda_\ell$ of the matrix $C$,
\begin{equation}
    S_A = - \sum_{\ell} (\lambda_\ell \ln{\lambda_\ell} + (1-\lambda_\ell) \ln{(1-\lambda_\ell)}).
\end{equation}

We shall restrict ourselves to the case where $A \subset X_{2d}$ is composed of the first $L$ neighborhoods 
of a given vertex $v_0 \in X_{2d}$, i.e.
\begin{equation}\label{eq:ball}
    \pi_A = \sum_{i = 0}^L E_i^*, \quad E_i^* = \sum_{\substack{ v \in X_{2d} \text{ s.t.}\\ \text{dist}(v,v_0) = i}} \ket{v}\bra{v}.
\end{equation}
Such subsystems are balls centered on an arbitrary point, and are natural generalizations of single intervals in one-dimensional systems. Moreover, the symmetry of region $A$ shall allow us to perform exact calculations through the use of dimensional reduction.

Since folded cubes are vertex-transitive, we can choose $v_0 = (0,0,\dots, 0)$ without loss of generality. The number of sites in the $i$-th neighborhood of $v_0$ is given by
\begin{equation}
    \text{tr}(E_i^*) = \binom{2d}{i} + \binom{2d}{2d+1 - i},
\end{equation}
and the computation of $S_A$ thus reduces to the diagonalization of a square matrix $C$ of dimension $|A| = \text{tr}(\pi_A)$,
\begin{equation}
    \text{tr}(\pi_A) = 1 +  \sum_{i=1}^L\left(\binom{2d}{i} + \binom{2d}{2d+1 - i}\right).
\end{equation}

\subsection{Dimensional reduction and $\mathfrak{so}(3)_{-1}$ }
The computation of $S_A$ can be further simplified by an explicit block-diagonalization of $C$. Indeed, the truncated correlation matrix is part of the matrix algebra $\mathcal{T}$ generated by the projectors $E_k$ onto eigenspaces of the adjacency matrix and the projectors $E_i^*$ onto neighborhoods of the vertex $v_0 = (0,0,\dots, 0)$,
\begin{equation}
    C = \sum_{i,j \leqslant L} \sum_{k \leqslant K} E_i^* \left(E_{2k} + E_{2k+1}\right) E_j^*.
\end{equation}

Folded cubes are distance-regular graphs that satisfy the $Q$-polynomial property. Consequently, the algebra $\mathcal{T}$, known as the Terwilliger algebra of folded cubes, possesses interesting properties \cite{t1,t2,t3}. It is semi-simple and equivalent to the algebra generated by the adjacency matrix $\A$ and the dual adjacency matrix $\A^*$,
\begin{equation}
    \mathcal{T} = \langle \A, \A^*\rangle,
\end{equation}
where $\A^*$ is the diagonal matrix whose entries are given by
\begin{equation}
    \bra{v}\A^*\ket{v} = 2^{2d}\bra{v}\left(E_{0} + E_{1}\right)\ket{v_0} = (-1)^{\text{wt}(v)}(2d +1 - 2\text{wt}(v)),
\end{equation}
where the weight function $\text{wt}$ counts the number of $1$ in a binary sequence,
\begin{equation}
    \text{wt}(v) = \sum_{i = 1}^{2d} v_i.
\end{equation}

The dual adjacency matrix can be expressed in the basis \eqref{bas1} in terms of Pauli matrices as
\begin{equation}
    \A^* =  \left(\sum_{n = 1}^{2d} \underbrace{\sigma_z \otimes \sigma_z \otimes ... \otimes \sigma_z}_{n-1 \text{ times}} \otimes \ I \otimes \underbrace{\sigma_z \otimes ... \otimes \sigma_z}_{2d-n \text{ times}}\right) + \underbrace{\ \sigma_z \otimes \ \sigma_z \otimes ... \otimes \ \sigma_z}_{2d \text{ times}}.
    \label{adjdu}
\end{equation}

Using the expressions \eqref{adj} and \eqref{adjdu}, one can show that the matrices
\begin{equation}
    K_1 =(-1)^d \A/2, \quad K_2 = \{\A, \A^*\}/4, \quad K_3 = (-1)^d \A^*/2
\end{equation}
verify the {following} defining relations of the algebra $\mathfrak{so}(3)_{-1}$,
\begin{equation}
    \{K_1,K_2\} = K_3, \quad  \{K_2,K_3\} = K_1, \quad  \{K_3,K_1\} = K_2.
\end{equation}
The algebra $\mathfrak{so}(3)_{-1}$ amounts to the anti-commutator version of the Lie algebra $\mathfrak{so}(3)$, and it has he following Casimir operator,
\begin{equation}
    \boldsymbol{K}^2 = K_1^2 + K_2^2 + K_3^2, \quad [\boldsymbol{K}^2, K_i] = 0.
\end{equation}

From this identification, one gets that the vector space $\mathbb{C}^{2^{2d}}$ onto which the adjacency matrices $\A$, $\A^*$ and truncated correlation matrix $C$ act is a $\mathfrak{so}(3)_{-1}$-module. Using the standard representation theory of $\mathfrak{so}(3)_{-1}$ \cite{genest}, this module can be decomposed into its irreducible components $\mathcal{V}_{j,r}$:
\begin{equation}
    \mathbb{C}^{2^{2d}} = \bigoplus_{j = 0}^{d} \bigoplus_{r = 1}^{D_j} \mathcal{V}_{j,r}, \quad D_j = \left\{
	\begin{array}{ll}
		\frac{2j + 1}{2d+1} \binom{2d+1}{d-j} + \frac{2j + 3}{2d+1} \binom{2d+1}{d-j-1} & \mbox{if } j \in \{0,1,2,\dots, d-1\},\\
		1 & \mbox{if } j = d,
	\end{array}
\right.
\end{equation}
where $\mathcal{V}_{j,r}$ is a subspace of $\mathbb{C}^{2^{2d}}$ spanned by vectors $\ket{j,r,n}_3$,
\begin{equation}
   \mathcal{V}_{j,r} = \text{span}\{ \ket{j,r,n}_3 \ | \ n \in \{0,1,\dots,j\}\}.
\end{equation}

The matrices $K_1$ and $K_3$ act on these vectors respectively as tridiagonal and diagonal matrices,
\begin{equation}
\begin{split}
       K_1 \ket{j,r,n}_3 &= \sqrt{\frac{(j+n+2)(j-n)}{4}} \ket{j,r,n+1}_3 + \delta_{n,0}\left(\frac{j+1}{2}\right)\ket{j,r,n}_3 \\
       &+ (1 - \delta_{n,0})\sqrt{\frac{(j+n+1)(j+1-n)}{4}} \ket{j,r,n-1}_3, \\[.3cm]
        K_3 \ket{j,r,n}_3 &= (-1)^n \left( n + \frac{1}{2}\right) \ket{j,r,n}_3.
\end{split}
\label{lequ}
\end{equation}
The Casimir $\boldsymbol{K}^2$ also acts on $\mathcal{V}_{j,r}$ as a multiple of the identity,
\begin{equation}
    \boldsymbol{K}^2\ket{j,r,n}_3 = ((j+1)^2 - 1/4)\ket{j,r,n}_3. 
\end{equation}
The matrix $K_3$ being diagonal in this basis, one finds that the projectors $E_i^*$ have a simple action,
\begin{equation}
    E_i^* \ket{j,r,n}_3 =\delta_{n, d-i}\ket{j,r,n}_3.
    \label{proy1}
\end{equation}

The representation theory of  $\mathfrak{so}(3)_{-1}$ further guaranties the existence of an alternative basis for the modules $\mathcal{V}_{j,r}$, such that the roles of $K_1$ and $K_3$ are inverted. In other words, we have
\begin{equation}
     \mathcal{V}_{j,r} = \text{span}\{ \ket{j,r,k}_1\ | \ k \in \{0,1,\dots,j\}\},
\end{equation}
where the action of $K_1$ and $K_3$ on $\ket{j,r,k}_1$ is given by
\begin{equation}
\begin{split}
       K_3 \ket{j,r,k}_1 &= \sqrt{\frac{(j+k+2)(j-k)}{4}} \ket{j,r,k+1}_1 + \delta_{k,0}\left(\frac{j+1}{2}\right)\ket{j,r,k}_1 \\
       &+ (1 - \delta_{k,0})\sqrt{\frac{(j+k+1)(j+1-k)}{4}} \ket{j,r,k-1}_1, \\[.3cm]
        K_1 \ket{j,r,k}_1 &= (-1)^k \left( k + \frac{1}{2}\right) \ket{j,r,k}_1.
\end{split}
\label{requ}
\end{equation}
In this second basis, one finds a simple action of the projectors $E_{2k} + E_{2k+1}$ onto eigenspaces of $K_1$,
\begin{equation}
\left(E_{2k} +E_{2k+1}\right) \ket{j,r,k'}_1 = \left(\delta_{2k -d, k'} + \delta_{d-2k-1, k'}  \right)\ket{j,r,k'}_1.
\end{equation}

The overlaps $ Q_{k,n} := \prescript{}{1}{\bra{j,r,k}\ket{j,r,n}}_{3} $ between these two bases of the submodule $\mathcal{V}_{j,r}$ can be computed explicitly. Indeed, equating the action of the Hermitian operator $K_3$ on the left and on the right in $\prescript{}{1}{\bra{j,r,k}K_3\ket{j,r,n}}_{3}$ and using \eqref{lequ} and \eqref{requ} yield the three term recurrence relation
\begin{equation}
\begin{split}
       (-1)^{n}(n + 1/2) Q_{n,k} &= \sqrt{\frac{(j+k+2)(j-k)}{4}} Q_{n,k+1}+ \delta_{k,0}\left(\frac{j+1}{2}\right)Q_{n,k} \\
       &+ (1 - \delta_{k,0})\sqrt{\frac{(j+k+1)(j+1-k)}{4}}Q_{n,k-1}.
\end{split}
\label{rr1}
\end{equation}
This recurrence is solved by anti-Krawtchouk polynomials $\hat{P}_n(x_k)$ evaluated on the grid $x_k = (-1)^k (k + 1/2)$ and modulated by appropriate weights $\Omega_k$ and normalisation functions $\Phi_n$,
\begin{equation}
     Q_{k,n} = \sqrt{\frac{\Omega_k}{\Phi_n}}\hat{P}_n(x_k).
\end{equation}
Explicit expressions for these functions are provided in App.~\ref{appA}. 

The overlaps $Q_{k,n}$ provide analytical expressions for the entries of the truncated correlation matrix in the basis of vectors $\ket{j,r,n}_3$,
\begin{equation}
    \prescript{}{3}{\bra{j,r,n}C\ket{j',r',n'}}_{3} = \delta_{j,j'}\delta_{r,r'} \left(\sum_{k = \left \lceil{\frac{d-j-1}{2}}\right \rceil}^{\text{min}\{K, \left \lceil{d/2}\right \rceil - 1 \}}Q_{d-2k -1,n} Q_{d-2k-1,n'} + \sum_{k = \left \lceil{d/2}\right \rceil}^{\text{min}\{K, \left \lfloor{\frac{d+j}{2}}\right \rfloor \}} Q_{2k-d,n} Q_{2k-d,n'} \right).
    \label{ent1}
\end{equation}

In this basis, the truncated correlation matrix $C$ exhibits a block-diagonal structure, where each submatrix $C|_{\mathcal{V}_{j,r}}$ depends solely on the value of $j$ and is independent of $r$. This property arises from the isomorphism between the submodules $\mathcal{V}_{j,r}$ corresponding to different {values of $r$.} 
Exploiting this block-diagonalization, we  derive the following formula for the entanglement entropy $S_A$,
\begin{equation}
    S_A = \sum_{j = 0}^d D_j S(j),
    \label{eef}
\end{equation}
where the terms $S(j)$ are given by
\begin{equation}\label{eq:Sjlambda}
    S(j) = -\sum_{\ell} (\lambda_{j,\ell} \ln \lambda_{j,\ell} + (1 - \lambda_{j,\ell})  \ln(1-\lambda_{j,\ell}))
\end{equation}
and $\lambda_{j,\ell}$ are the eigenvalues of the submatrix $C|_{\mathcal{V}_{j,r}}$ restricted to the irreducible subspace $\mathcal{V}_{j,r}$ and with entries given by \eqref{ent1}.

\subsection{Anti-Krawtchouk chains and $S(j)$}
The coefficients $S(j)$ in \eqref{eef} for the entanglement entropy $S_A$ can be interpreted as the entanglement entropy of inhomogeneous one-dimensional free-fermion systems. Indeed, in terms of the fermion operators
\begin{equation}
    b_{j,r,n} = \sum_{v \in X_{2d}}  \prescript{}{3}{\bra{j,r,n}\ket{v}} c_v, \quad  b_{j,r,n}^\dagger = \sum_{v \in X_{2d}} \bra{v}\ket{j,r,n}_3 c_v^\dagger,
\end{equation}
one can rewrite the Hamiltonian $\mathcal{H}$ in \eqref{eq:H} as a sum of Hamiltonians $\mathcal{H}_{j,r}$ acting on independent degrees of freedom,
\begin{equation}
    \mathcal{H} = \sum_{j = 0}^d \sum_{r = 1}^{D_j} \mathcal{H}_{j,r}
    \label{decomp}
\end{equation}
where $\mathcal{H}_{j,r}$ describes free fermions on an inhomogeneous chain of length $j+1$, governed by the Hamiltonian
\begin{equation}
    \mathcal{H}_{j,r} = (j+1)b_{j,r,0}^\dagger b_{j,r,0} + \sum_{n= 0}^{j-1} \sqrt{(j+n+2)(j-n)} (b_{j,r,n+1}^\dagger b_{j,r,n} +b_{j,r,n}^\dagger b_{j,r,n+1} ).
\end{equation}

Since these Hamiltonians can be diagonalized in terms of anti-Krawtchouk polynomials, these are referred to as \textit{anti-Krawtchouk chains}. From the point of view of these chains, the ground state $|\Psi_0\rrangle$ is expressed as the tensor product of ground states $|\Psi_0\rrangle_{j,r}$ of each chain in the decomposition \eqref{decomp},
\begin{equation}
    |\Psi_0\rrangle = \bigotimes_{j=0}^d\bigotimes_{r=1}^{D_j} |\Psi_0\rrangle_{j,r},
    \label{prol1}
\end{equation}
and the correlation matrix decomposes as 
\begin{equation}
    \hat{C} = \bigoplus_{j = 0}^d \bigoplus_{r = 1}^{D_j} \hat{C}|_{\mathcal{V}_{j,r}} 
    \label{d11}
\end{equation}
where $\hat{C}|_{\mathcal{V}_{j,r}} $ is the correlation matrix in the ground state of $\mathcal{H}_{j,r}$. Given the simple action \eqref{proy1} of $E_i^*$ on each module $\mathcal{V}_{j,r}$, we also have that
\begin{equation}
    \pi_A = \bigoplus_{j = 0}^d \bigoplus_{r = 1}^{D_j} \pi_A|_{\mathcal{V}_{j,r}}
    \label{prol2}
\end{equation}
where $\pi_A|_{\mathcal{V}_{j,r}}$ is the projector onto the last $L-d+j +1$ sites of the anti-Krawtchouk chain $j,r$,
\begin{equation}
    \pi_A|_{\mathcal{V}_{j,r}} = \sum_{n = d-L}^{j} \ket{j,r,n}_3 \prescript{}{3}{\bra{j,r,n}}.
    \label{d22}
\end{equation}

Using \eqref{d11} and \eqref{d22}, one recovers a block-diagonalization of the truncated correlation matrix $C$,
\begin{equation}
    {C} = \bigoplus_{j = 0}^d \bigoplus_{r = 1}^{D_j} \pi_A\hat{C}\pi_A |_{\mathcal{V}_{j,r}} =  \bigoplus_{j = 0}^d \bigoplus_{r = 1}^{D_j} {C}|_{\mathcal{V}_{j,r}}
    \label{d1}
\end{equation}
where $C|_{\mathcal{V}_{j,r}}$ is now interpreted as the truncated correlation matrix associated to the last $L - d + j +1$ sites of the ground state of  $\mathcal{H}_{j,r}$. The coefficient $S(j)$ in \eqref{eef} thus corresponds to the entanglement entropy contribution coming from the intersection of the region $A$ and the chain associated to the module $\mathcal{V}_{j,r}$. 

\subsection{Numerical investigation of $S(j)$}

{In the following, we investigate the scaling of $S(j)$ as a function of $j$. The entanglement properties of inhomogeneous free-fermion chains solved by orthogonal polynomials have been intensely studied recently \cite{Crampe:2019upj,FA20,FA21,Crampe:2021,bernard2022entanglement,bernard2022computation}.} 

\begin{figure}
    \centering
\includegraphics[scale = 1]{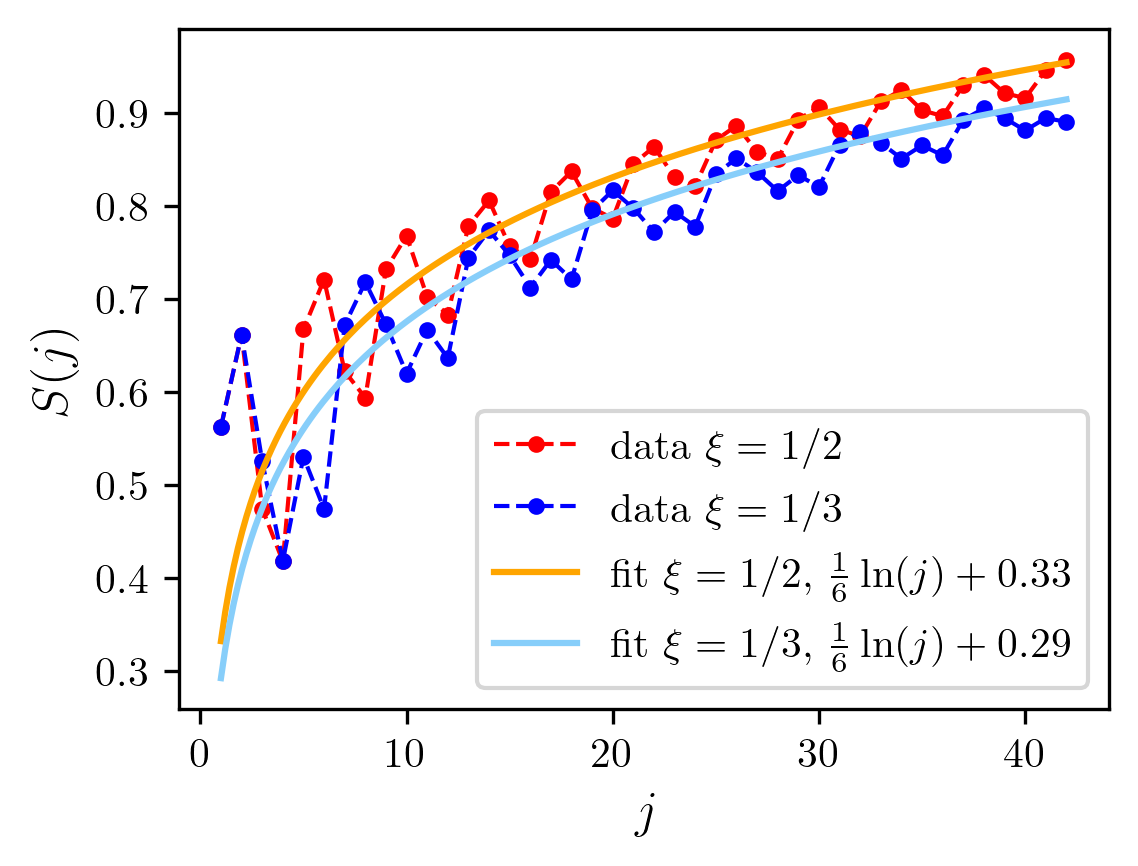}
    \caption{Scaling of the entanglement entropy $S(j)$ for half anti-Krawtchouk chains at half-filling. Oscillations are due to sub-leading terms and vanish with large $j$. Solid lines were obtained by fitting \eqref{sc1} with the unknown coefficient $a_1(\kappa,\xi)$.}
    \label{fig:sc1}
\end{figure}

We fix the ratios $\kappa := K/d$ and $\xi :=(d-L)/(j+1)$ and use \eqref{eq:Sjlambda} to compute $S(j)$ via exact numerical diagonalization of the chopped correlation matrix \eqref{ent1}. Physically, the ratio $\kappa$ corresponds to the filling fraction, whereas $\xi$ is (one minus) the aspect ratio, namely the ratio between the size $j+1$ of the chain, and the length $d-L$ of the complement of the intersection between $A$ and the chain. We present our numerical results for chains at half-filling, $\kappa=1/2$, in Fig.~\ref{fig:sc1}. We find a scaling of the form

\begin{equation}
    S(j) = \frac{1}{6}\ln(j) + a_1(\kappa,\xi) +  o(1)
    \label{sc1}
\end{equation}
in the limit of large $j$. This corresponds to a logarithmic violation of the area law in one dimension \cite{CC04}, which is typical for one-dimensional free-fermion models described by an underlying CFT with central charge $c=1$. The scaling \eqref{sc1} thus suggests that the anti-Krawtchouk chain is described by a CFT in curved space \cite{DSVC17} with $c=1$, similarly to the Krawtchouk chain \cite{FA21,bernard2022entanglement}. 

The presence of oscillations in Fig.~\ref{fig:sc1} can be attributed to sub-leading terms that have not been fully characterized in the present analysis. Similar oscillations have been observed in Krawtchouk chains and a conjecture regarding the sub-leading terms was proposed in \cite{bernard2022entanglement}.

\begin{figure}
\begin{subfigure}{.5\textwidth}
  \centering
  \includegraphics[width=.9\linewidth]{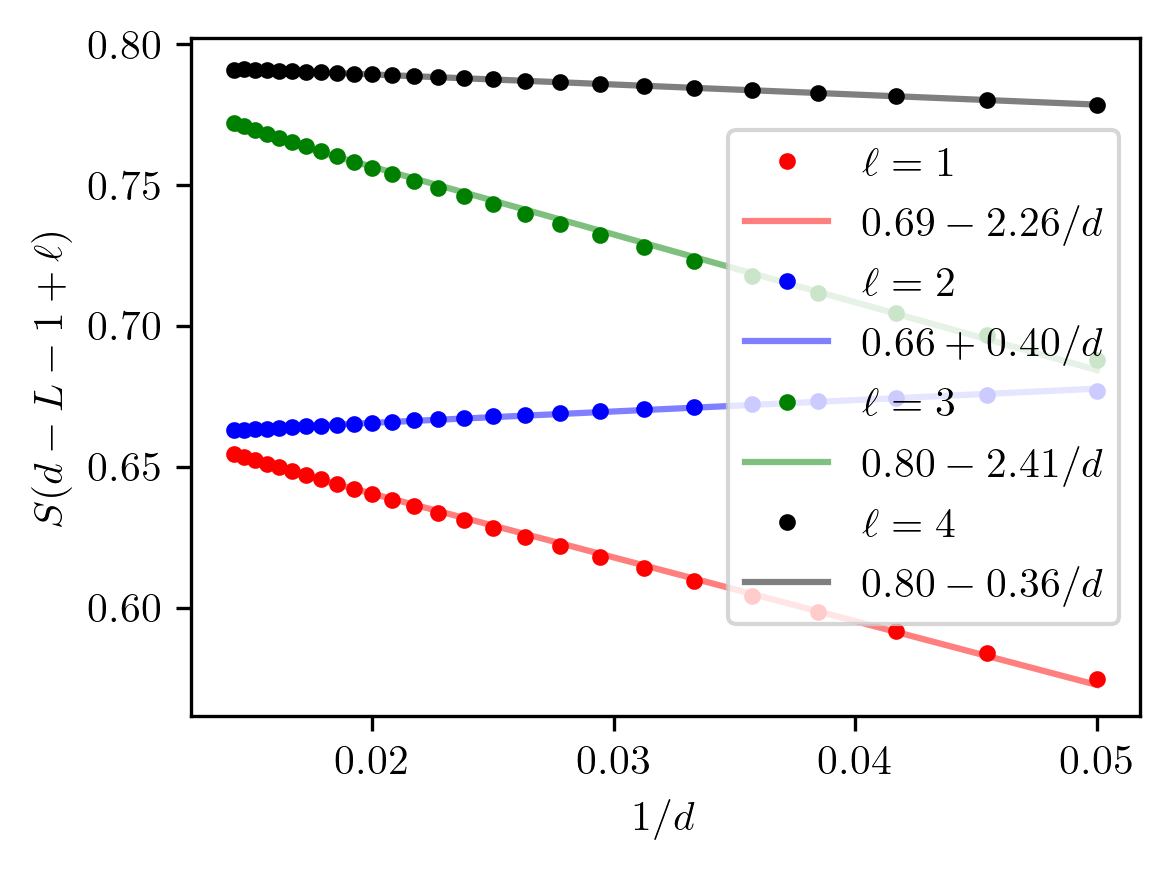}
  \label{fig:sfig1}
\end{subfigure}%
\begin{subfigure}{.5\textwidth}
  \centering
  \includegraphics[width=.9\linewidth]{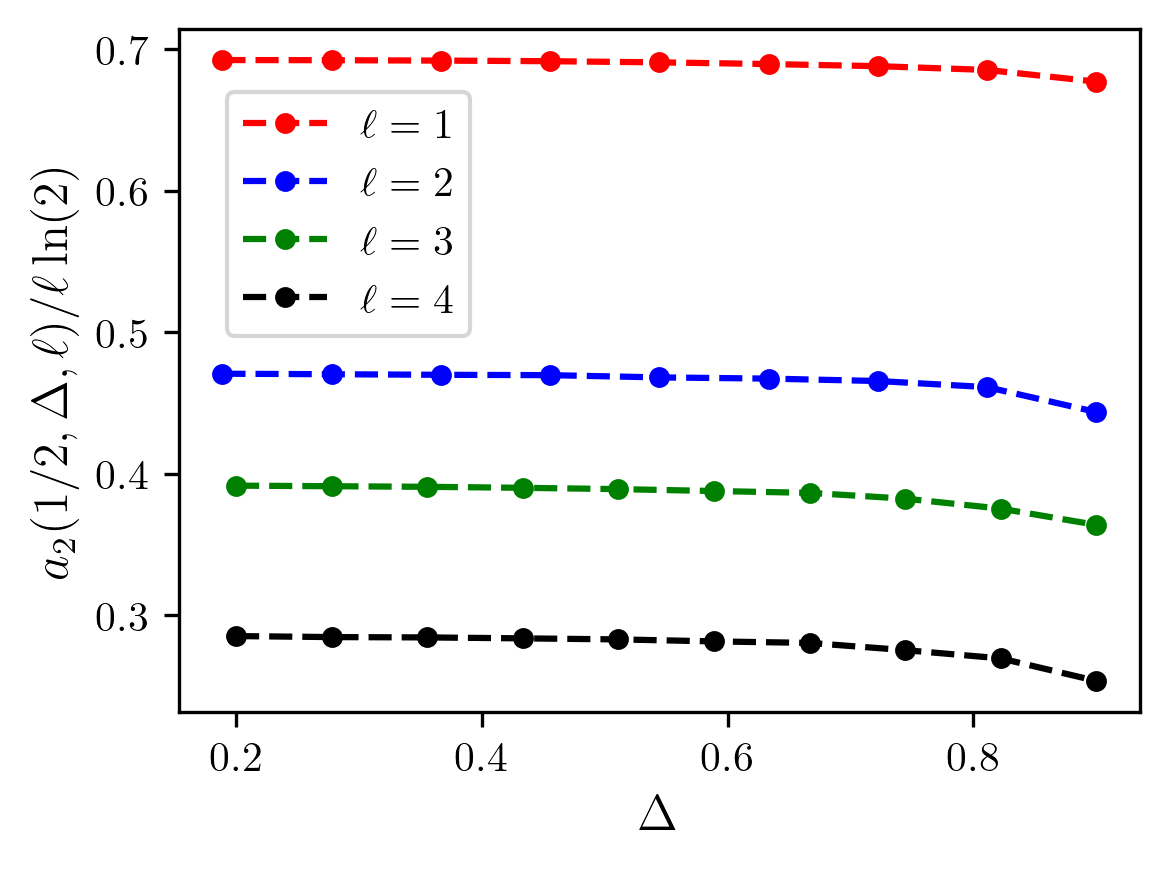}
  \label{fig:sfig2}
\end{subfigure}
\caption{Entanglement entropy of anti-Krawtchouk chains in the large-diameter limit with fixed filling ratio $\kappa$, aspect ratio $\Delta = L/d$ 
and subsystem size $\ell$. The left figure illustrates the convergence of $S(d-L-1 + \ell)$ to a value $a_2(\kappa,\Delta,\ell)$ of magnitude near $\ln(2)$ at large diameter $d$ and $\kappa = \Delta = 1/2$. The right figure presents the value of the ratio $a_2(\kappa,\Delta,\ell)/\ell \ln(2)$ at $\kappa = 1/2$ for various $\Delta$ and $\ell$. }
\label{fig:entell}
\end{figure}

For a fixed filling ratio $\kappa$, a fixed ratio $\Delta := L/d$ and a fixed subsystem size $\ell = L-d+j+1$ in the chain, the entanglement entropy rather converges at large diameter $d$ to a constant value,
\begin{equation}
    S(d-L-1 + \ell) = a_2(\kappa, \Delta, \ell) + o(1) \leqslant \ell \ln(2).
    \label{sc2p}
\end{equation}
Here, the bound on the entanglement entropy is determined by its value for a maximally entangled state given a subsystem of size $\ell$. Numerical analysis verifies that the magnitude of $a_2(\kappa, \Delta, \ell)$ remains close to $\ln(2)$ even for $\ell \neq 1$, indicating as expected that most of the entanglement originates from a highly entangled state at the boundary, see Fig. \ref{fig:entell}.

\subsection{Large diameter limit of the folded cube}
Let us fix the ratio $\Delta := L/d$, where $L$ is the diameter of the region $A$ in the folded cube and $d$ is the diameter of the graph. We shall now consider how the entanglement entropy scales in the limit of large diameter $d$ at half-filling $K = d/2$. From numerical tests and the scaling given by \eqref{sc1} and \eqref{sc2p}, we find that the behavior of $S(j)$ is captured by
\begin{equation}
    S(j) \sim
\left\{
	\begin{array}{ll}
 \frac{1}{6} \ln{(d)} + O(1) & \mbox{if }  1-\Delta < j/d, \\[.2cm]
		0  & \mbox{if } 1-\Delta > j/d, \\[.2cm]
        a_2(\kappa, \Delta, \ell) & \mbox{if }  1-\Delta  \approx j/d.
	\end{array}
\right.
\label{sc2}
\end{equation}
In the first situation, $1-\Delta < j/d$, the scaling behavior follows equation \eqref{sc1}. For $1-\Delta > j/d$, the contribution $S(j)$ arises from anti-Krawtchouk chains that do not intersect with the region $A$, resulting in a zero contribution. In the third regime, $1-\Delta \approx j/d$, the chains have a small intersection $\ell \ll d$ 
 with the region $A$ compared to their size of $j+1$; the region $A$ in these chains is predominantly composed of a highly entangled site on the boundary, leading to a contribution of $a_2(\kappa, \Delta, \ell) \leqslant\ell \ln(2)$ to $S(j)$. 
 
 In the limit of large diameter, Stirling's formula provides an asymptotic expression for the degeneracy,
\begin{equation}
    D_{j} \sim  \frac{j4^{d+1}}{d\sqrt{d \pi}} e^{- j^2 /d}.
\end{equation}
In particular, one notes that the degeneracy reaches a peak at $j  \sim \sqrt{d/2}$ and then gets exponentially small as~$j$ increases. It follows that the largest contribution to the entanglement entropy is coming from terms $D_j S(j)$ in equation \eqref{eef} for which $j/d$ is small but larger than $1 - \Delta$. This corresponds to the third regime $1-\Delta  \approx j/d$ and justifies the following expression for the scaling of $S_A$:
\begin{equation}
\begin{split}
    S_A  &= D_{d-L} \sum_{\ell = 1}^{L+1} \frac{D_{d-L-1 + \ell}}{D_{d-L}} S(d-L -1 + \ell) \\
    & \sim D_{d-L} \sum_{\ell = 1}^{L+1} e^{- 2 (1-\Delta) (\ell - 1)} S(d-L  -1+\ell).
\end{split}
    \label{sc3}
\end{equation}

Using the bound \eqref{sc2p} on $S(d-L -1 +\ell)$, we find that the entanglement entropy at large $d$ and fixed $\Delta$ and $\kappa$, is bounded by a strict area law, with no logarithmic enhancement,
\begin{equation}
\begin{split}
    S_A   \leqslant & D_{d-L} \sum_{\ell = 1}^{L+1} e^{- 2 (1-\Delta) (\ell - 1)} \ell \ln{2} 
    \\
    & \sim  D_{d-L} \ln(2) \frac{e^{4(1- \Delta)}}{(e^{2(1- \Delta)} - 1)^2}\\
    &\sim |\partial A|\ln(2) \frac{e^{2(1-\Delta)}}{e^{2(1-\Delta)} - 1}.
\end{split}
    \label{sc3p}
\end{equation}
Here, we used the following approximation for the area of the boundary $\partial A$,
\begin{equation}
    |\partial A| = \sum_{\ell = 1}^{L+1} D_{d-L-1+\ell } \sim D_{d-L} \frac{e^{2(1-\Delta)}}{e^{2(1-\Delta)} - 1},
\end{equation}
which also corresponds to the number of anti-Krawtchouk chains intersecting the region $A$.
\begin{figure}
    \centering
    \includegraphics[scale = 1]{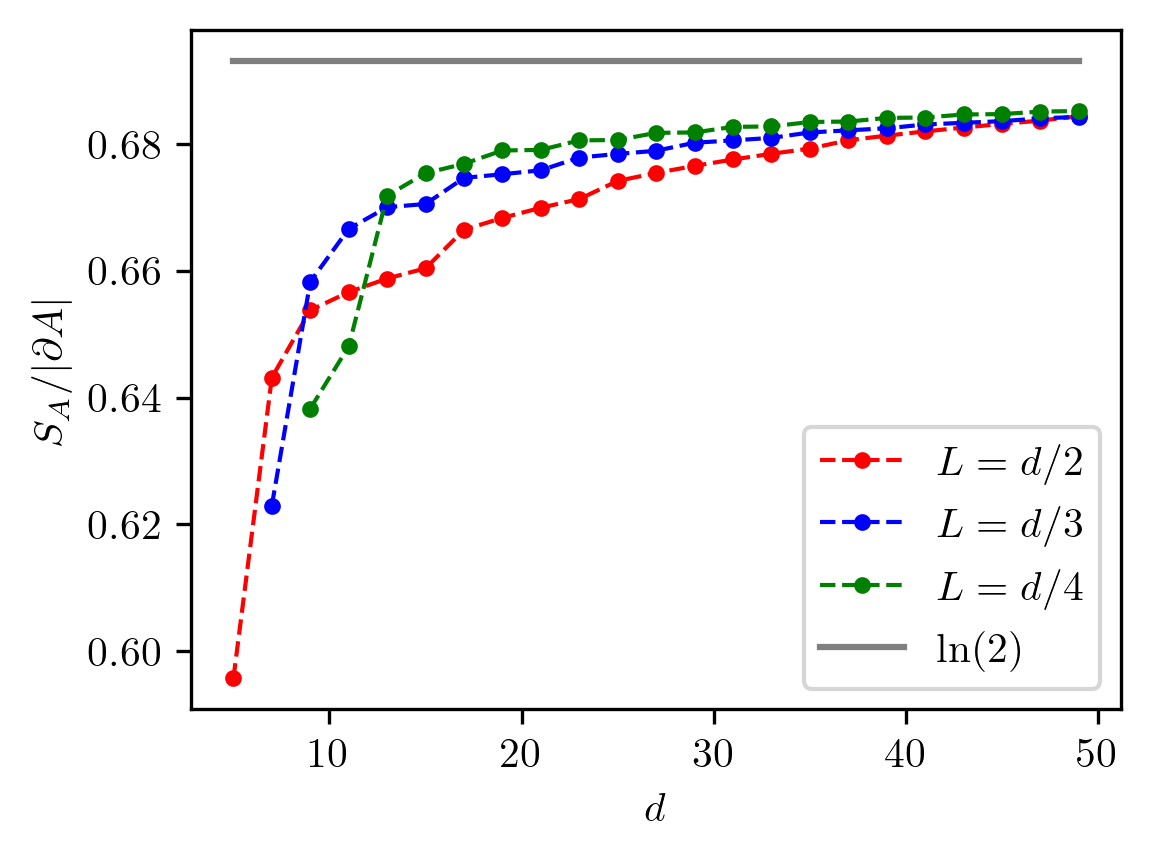}
    \caption{Ratio of the entanglement entropy $S_A$ over the boundary area $|\partial A|$ for region $A$ composed of the first $L$ neighborhoods of a vertex in a folded cube, at half filling $\kappa = 1/2$. The entanglement entropy $S_A$ is obtained by numerical diagonalization of the truncated {correlation} matrix $C$. In the large-diameter $d$ limit, the ratio $S_A/|\partial A|$ converges near $\ln(2)$ and shows no logarithmic enhancement.  }
    \label{ak10}
\end{figure}
Furthermore, since $S(d-L-1 +\ell)$ converges at large $d$ to a value near $\ln(2)$ (as shown in Fig.~\ref{fig:entell}), a good estimate of the magnitude of $S_A$ is provided by $ S_A \approx |\partial A| \ln(2)$, as illustrated in Fig.~\ref{ak10}.

Equation \eqref{sc3p} exhibits a strict area law, which is atypical for critical free-fermion systems that usually display logarithmic enhancements of the area law in the scaling of the entanglement entropy. This behavior aligns with observations made for entanglement entropy in free fermions on high-dimensional structures such as Hamming and Johnson graphs \cite{bernard2023entanglement, bernard2021entanglement, parez2022multipartite}. 

The underlying cause can be attributed to the high-dimensional geometry of these graphs, wherein the majority of degrees of freedom on the boundary of region $A$ do not fully perceive the system's entire diameter. To be more specific, these graphs can be effectively decomposed into a combination of one-dimensional systems, most of which have only a small fraction of their degrees of freedom originating from the region $A$ and its boundary $\partial A$. From the perspective of the free-fermion chains within this decomposition, the process of taking the large-diameter limit of the graph does not correspond to a thermodynamic limit. As a result, their contribution to the entanglement entropy does not give rise to logarithmic enhancements. The area law for entanglement in ground states is a highly local property, caused by correlations of degrees of freedom close to the boundary. Therefore, we expect our findings to hold at leading order for arbitrary regions with volume, not only for the symmetric balls defined in \eqref{eq:ball}. However, our analytical methods are not applicable in such cases, and numerical calculations are cumbersome due to the exponentially large dimension of the Hilbert space.

\section{Entanglement Hamiltonian and Heun operator}
\label{s4}
While the entanglement entropy $S_A$ gives insight into the entanglement properties of the ground state, a more complete picture lays in the reduced density matrix $\rho_A$ and the entanglement Hamiltonian $\mathcal{H}_{\text{ent}}$.
Since the ground state is the product of ground states of anti-Krawtchouk chains \eqref{prol1} onto which the projector $\pi_A$ acts simply, the reduced density matrix can be decomposed as
\begin{equation}
    \rho_A = \bigotimes_{j =0}^d \bigotimes_{r = 1}^{D_j} \rho_A(j,r), \quad \rho_A(j,r) = \text{tr}_B|\Psi_0\rrangle_{j,r} \llangle \Psi_0|_{j,r}.
\end{equation}

The entanglement Hamiltonian can further be expressed as a sum of quadratic operators $\mathcal{H}_{\text{ent}}(j,r)$ acting on individual chains,
\begin{equation}
    \mathcal{H}_{\text{ent}} = \sum_{j=0}^d \sum_{r=1}^{D_j}  \mathcal{H}_{\text{ent}}(j,r), \quad \rho_A(j,r) = \frac{e^{-\mathcal{H}_{\text{ent}}(j,r)}}{\text{tr}_A e^{-\mathcal{H}_{\text{ent}}(j,r)}},
\end{equation}
\begin{equation}
    \mathcal{H}_{\text{ent}}(j,r) = \sum_{n,m = d-L}^j  \prescript{}{3}{\bra{j,r,n}h\ket{j,r,m}}_{3} b^\dagger_{j,r,n} b_{j,r,m},
\end{equation}
{where the matrix $h$ is defined in \eqref{enthamil}.}

The characterization of the reduced density matrix $\rho_A$ thus amounts to describing the entanglement Hamiltonian for each anti-Krawtchouk chain.  To streamline the analysis, we will focus on a single chain, denoted as $j,r$, or a single module $\mathcal{V}_{j,r}$ at a time and use the following abbreviated notation:
\begin{equation}
    \varrho_A := \rho_A(j,r), \quad \mathfrak{C} := C|_{\mathcal{V}_{j,r}}, \quad \hat{\mathfrak{C}} := \hat{C}|_{\mathcal{V}_{j,r}},  \quad \mathfrak{h} := h|_{\mathcal{V}_{j,r}}, \quad \mathfrak{K}_i := K_i|_{\mathcal{V}_{j,r}}
\end{equation}

\subsection{Commuting Heun operator and $\varrho_T(t_1,t_2)$}
In order to describe $\varrho_A$, we are interested in the identification of the matrix $\mathfrak{h}$. It can be done using the correlation matrix as an input in the formula $\mathfrak{h} = \ln\left((1-\mathfrak{C}) / \mathfrak{C}\right)$ \cite{peschel2003calculation}. This is straightforward but does not provide an explicit formula for the entries $\mathfrak{h}_{nm}$. Moreover, it can be numerically unstable due to the proximity of most eigenvalues of $\mathfrak{C}$ to $0$ and $1$. An alternative method is to use the fact that $\mathfrak{C}$ admits a simple commuting operator known as a generalized algebraic Heun operator $T$,
\begin{equation}
    T = \{\mathfrak{K}_1 - \mu , \mathfrak{K}_3^2 - \nu \},
\end{equation}
where the coefficients $\mu$ and $\nu$ are given by
\begin{equation}
    \mu = (-1)^d(2K - d + 3/2), \quad \nu = (d-L)^2 + 1/4.
\end{equation}

Indeed, one can check using the representation of $\mathfrak{K}_1$ and $\mathfrak{K}_3$ in the basis of vectors $\ket{j,r,n}_1$ that $[T,\hat{\mathfrak{C}}] =0$. Similarly, the basis of vectors $\ket{j,r,n}_3$ makes it straightforward to check that the Heun operator commutes with the projector onto region $A$, i.e. $[T, \pi_A]=0$. It then follows that,
\begin{equation}
    [T, \mathfrak{C}] = [T, \mathfrak{h}] = 0.
    \label{comi}
\end{equation}

A commuting Heun operator also exists for homogeneous free fermion chains and a wide range of inhomegenous models \cite{eisler2013free,eisler2017analytical,eisler2018properties,grunbaum2018algebraic,eisler2019continuum,Crampe:2021,bernard2021heun,bernard2022entanglement,bernard2022computation}. Understanding precisely the relation between these commuting tridiagonal matrices, correlation matrices and entanglement Hamiltonians has attracted some attention (notably in the homogeneous case \cite{eisler2013free,eisler2017analytical,eisler2018properties,eisler2019continuum}) but remains in general an open question. Our aim is to express $\mathfrak{h}$ as a sum of powers of $T$. More precisely, we are interested in the possibility of approximating $\mathfrak{h}$ as an affine transformation of the Heun operator.
The matrix $ {T}$ is irreducible tridiagonal in the basis $\{\ket{j,r,n}_3 : d-L \leqslant n \leqslant j\}$ and is thus non-degenerate on $\pi_A \mathcal{V}_{j,r}$. It further commutes with $\mathfrak{h}$ so they can be related on this subspace by the following sum,
\begin{equation}
    \mathfrak{h} = \sum_{i = 1}^{N_j} t_i { {T}}^{i-1},
    \label{ses1}
\end{equation}
where  $N_j = L-d+j+1$ is the dimension of the subspace $\pi_A \mathcal{V}_{j,r}$ and the coefficients $t_1, t_2 ,\dots t_{N_j}$ are fixed such that both sides of equation \eqref{ses1} have the same set of eigenvalues. Since we are examining the ground state of a local system, we anticipate that the dominant elements of $\mathfrak{h}$ correspond to the hopping terms between neighboring sites. Given that the Heun operator exclusively connects nearest neighbors, i.e.
\begin{equation}
    \prescript{}{3}{\bra{j,r,n}T\ket{j,r,m}}_{3}  \neq 0 \quad \Rightarrow \quad |n-m| \leqslant 1,
\end{equation}
it suggests that $\mathfrak{h}$ could be approximated to some extent by the first two powers of ${T}$,
\begin{equation}
    \mathfrak{h} \sim t_1 + t_2 {T},
    \label{app}
\end{equation}
where $t_1$ and $t_2$ are left to be determined. For example, this relation with $t_1=0$ and $t_2=-\pi L$ holds in the continuum limit of homegeneous one-dimensional chains at half filling \cite{eisler2019continuum}, where $L$ is the length of the interval. For general $t_1$ and $t_2$, one can define an Hamiltonian $\mathcal{H}_{T}$ and density matrix $\varrho_T$ as
\begin{equation}
\mathcal{H}_{T}(t_1, t_2) := \sum_{n,m = d-L}^{j} \left( t_1 \delta_{nm} +t_2 {{T}_{nm}} \right)b_{j,r,n}^\dagger b_{j,r,m},
\end{equation}
and
\begin{equation}
    \varrho_T(t_1,t_2) := \frac{e^{-{\mathcal{H}}_{T}(t_1,t_2)}}{\text{tr}_A e^{-{\mathcal{H}}_{T}(t_1,t_2)}}.
    \label{ddmm}
\end{equation}

A natural idea to determine the coefficients $t_i$ is to minimize the distance between $\mathfrak{h}$ and the expansion \eqref{ses1}. However, this method is not efficient when the number of parameters is greater than one \cite{BE23pc}. Our approach to fix the parameters $t_1$ and $t_2$ is to require that $\varrho_A$ and $ \rho_T(t_1,t_2)$ agree on the expectation value of observables. Specifically, these density matrices can be selected such that they coincide in the expected number of particles and the von Neumann entropy $S(j)$ in the anti-Krawtchouk chain $j$,$r$:
\begin{subequations}\label{eq:c1c2}
\begin{equation}
       S(j) =- \text{tr}(\varrho_A\ln\varrho_A)  = - \text{tr}(\varrho_T(t_1,t_2)\ln\varrho_T(t_1,t_2)),
       \label{c2}
\end{equation}
\begin{equation}
    \langle Q_A \rangle = \text{tr}_A( Q_A \varrho_A) =  \text{tr}_A( Q_A \varrho_T(t_1,t_2)),
    \label{c1}
\end{equation}
\end{subequations}
where $Q_A$ is the operator counting the number of particles in the intersection of the region $A$ with the chain $j,r$,
\begin{equation}
    Q_A = \sum_{n = d-L}^{j} b_{j,r,n}^\dagger b_{j,r,n}.
\end{equation}


\subsection{R\'enyi fidelities and R\'eyni entropies 
of $\varrho_A$ and $\varrho_T(t_1,t_2)$}

{In this section we compare the reduced density matrix $\varrho_A$ with the affine approximation $\varrho_T(t_1,t_2)$, where $t_1$ and $t_2$ are fixed by the constraints \eqref{eq:c1c2}. To achieve this, we compute their R\'enyi fidelities and their respective R\'enyi entropies.}

{Quantum} fidelities {quantify the resemblance between two quantum states \cite{U76, J94}.} Importantly, fidelities can be used to detect and characterize quantum phase transitions \cite{QSLZS06,ZB08,G10,DS11,SD13,hagendorf2017open,PMDR19,MDPL21,HP21}, similarly to the entanglement entropy. R\'enyi fidelities \cite{parez2022symmetry} were introduced recently as a generalization of Uhlmann-Jozsa fidelity \cite{U76,J94}.
They are defined for general density matrices $\rho$ and $\sigma$ as 
\begin{equation}
    F_n(\rho, \sigma) = \frac{\text{tr}\{(\rho \sigma)^n\}}{\sqrt{\text{tr}\{\rho^{2n}\}\text{tr}\{\sigma^{2n}\}}}.
    \label{forform}
\end{equation}
and they verify the following properties,
\begin{subequations}
\begin{equation}
    0 \leqslant F_n(\rho, \sigma) \leqslant 1,
\end{equation}
\begin{equation}
   F_n(\rho, \sigma) =1  \quad \Leftrightarrow \quad \rho = \sigma.
\end{equation}
\end{subequations}

In the case of Gaussian fermionic states, the formula \eqref{forform} for R\'eyni fidelities reduces to an expression in terms of the eigenvalues of the correlation matrices of $\rho$ and $\sigma$ \cite{parez2022symmetry}. Applying this result to the commuting states $\varrho_A$ and $ \varrho_T(t_1, t_2)$, one finds
\begin{equation}
    F_n(\varrho_T(t_1,t_2),\varrho_A ) = \det\left( \frac{\mathfrak{C}^n(I + e^{t_1 + t_2 T})^{-n} + (I-\mathfrak{C})^n(I + e^{-t_1 - t_2T})^{-n} }{\sqrt{\mathfrak{C}^{2n} + (I-\mathfrak{C})^{2n}}\sqrt{(I + e^{t_1 + t_2 T})^{-2n} + (I + e^{-t_1 - t_2 T})^{-2n}}} \right).
    \label{Rff}
\end{equation}
\begin{figure}
\centering
\includegraphics[scale = 1]{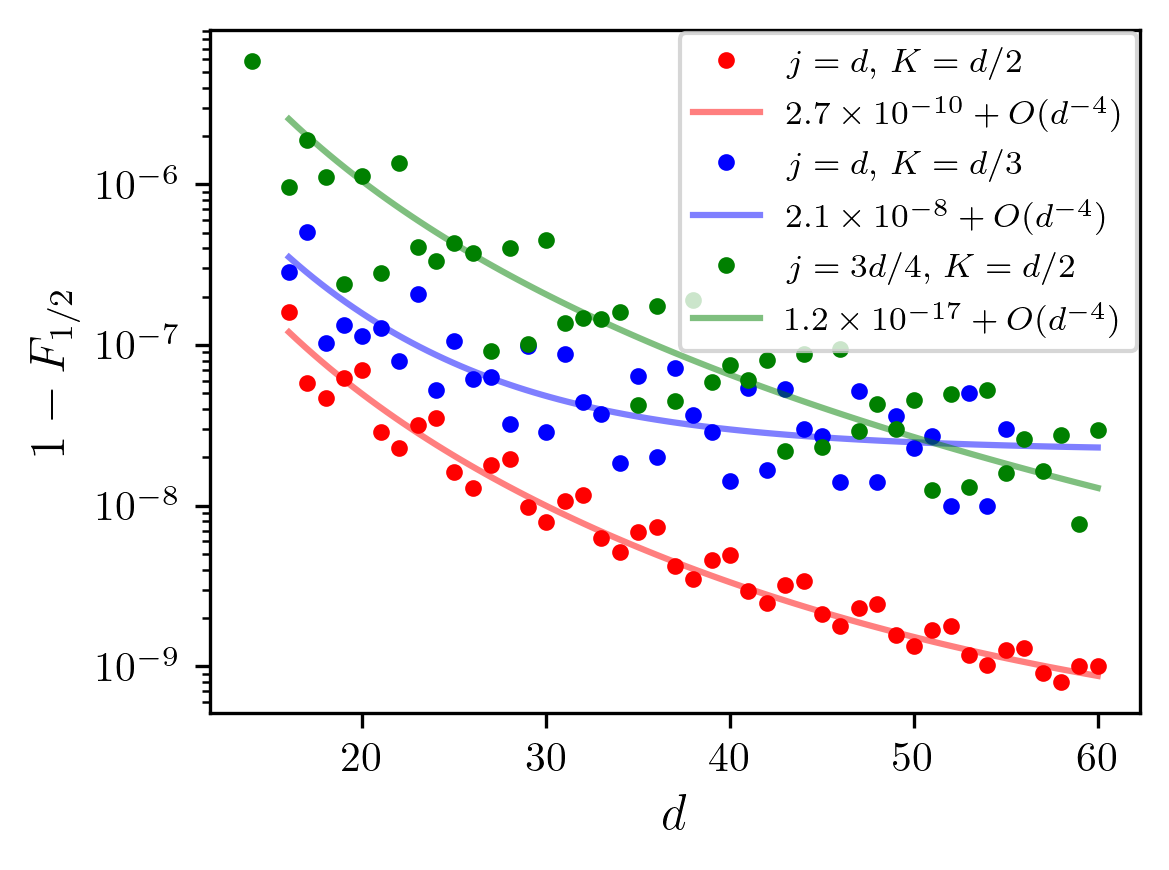}
\caption{R\'eyni fidelity $F_{1/2}$ between $\varrho_A$ and $\varrho_T(t_1,t_2)$, at $L = d/2$ and different parameters $j$, $d$ and $K$. It converges to a value near $1$ at large diameter.}
\label{AK14}
\end{figure}

The results of numerical computation of \eqref{Rff} for $n= 1/2$ and diameters of up to $d = 60$ are shown in Fig.~\ref{AK14}. In particular, $F_{1/2}$ seems to converge to a value very close to $1$ in the large-diameter limit. A similar behavior was also observed for general values of $n$.  The high fidelities between the two states demonstrate that $\varrho_T(t_1,t_2)$ captures the essence the reduced density matrix $\varrho_A$, confirming the validity of the linear approximation \eqref{app} and the local nature of the state $\varrho_A$. It also suggests that density matrices based on an affine transformation of Heun operators could offer a convenient approximation of the reduced density matrix of free-fermion ground states in other settings.

The proximity of the states $\varrho_A$ and $\varrho_T(t_1,t_2)$ is further visible in their respective  R\'eyni entropies. Indeed, one can consider their deviation $\delta S_\alpha$ defined as
\begin{equation}
\delta S_\alpha := \big|S_\alpha(\varrho_A) - S_\alpha(\varrho_T(t_1,t_2))\big|
\end{equation}
where $ S_\alpha(\rho)$ are the R\'enyi entropies,
\begin{equation}
    S_\alpha(\rho) = \frac{1}{1-\alpha}\ln \text{tr}(\rho^\alpha).
\end{equation}
Let us note that it is distinct from the \textit{relative entropy} between $\varrho_A$ and $\varrho_T(t_1,t_2)$, which also measures the proximity between two quantum states \cite{chuang} and would deserve an investigation of its own in this setting.

Using the known relation between $S_\alpha$ and the matrix $\mathfrak{h}$ \cite{carrasco2017duality}, $\delta S_\alpha$ can be expressed as
\begin{equation}
       \delta S_\alpha = \Bigg|\frac{1}{1-\alpha} \text{tr}\ln \left( \frac{\left(1 + e^ {\mathfrak{h}}\right)^{-\alpha}  + \left(1 + e^{-\mathfrak{h}}\right)^{-\alpha}}{\left(1 + e^ {t_1 + t_2{T}}\right)^{-\alpha}  + \left(1 + e^{-t_1 -t_2{T}}\right)^{-\alpha}}\right)\Bigg|.
\end{equation}
\begin{figure}
\begin{subfigure}{.5\textwidth}
  \centering
  \includegraphics[width=1.\linewidth]{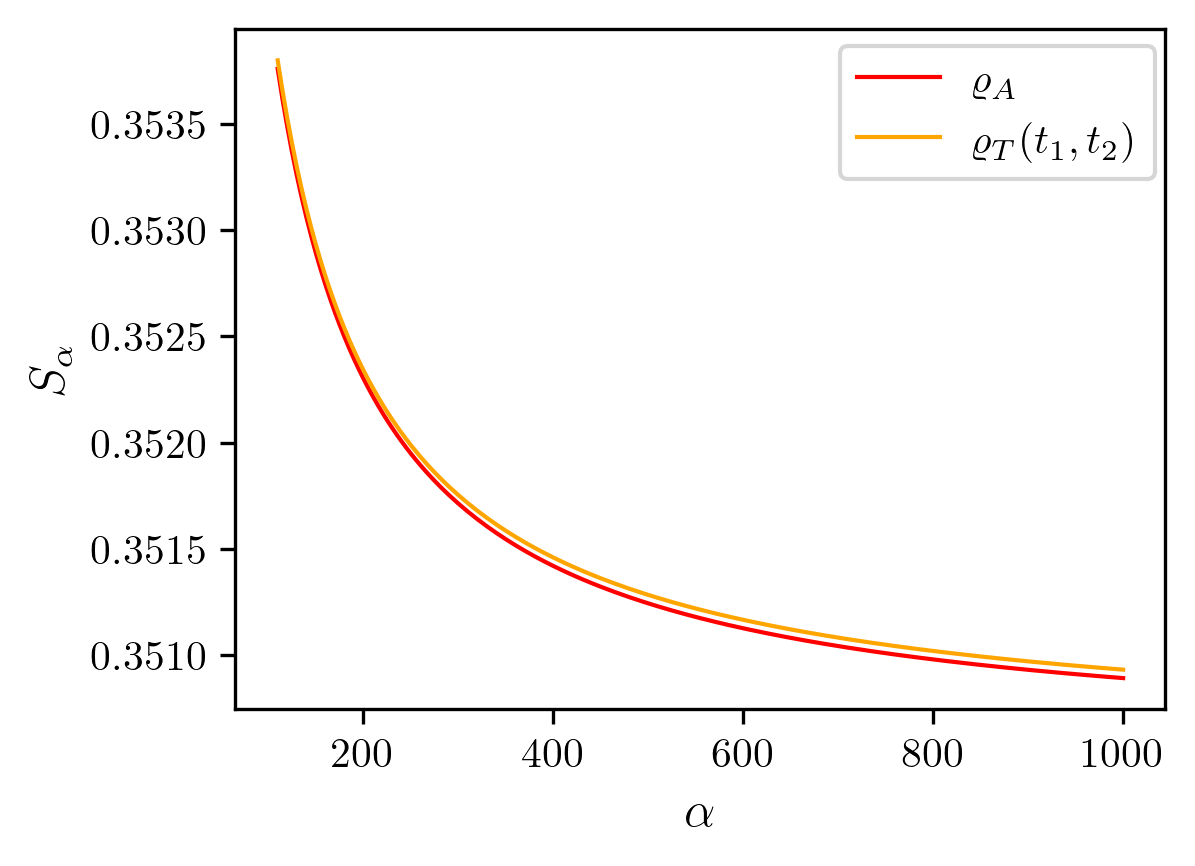}
  \caption{ $S_\alpha(\varrho_A)$ and $S_\alpha(\varrho_T(t_1,t_2))$ at $d= 40$.}
  \label{fig:sfig1}
\end{subfigure}%
\begin{subfigure}{.5\textwidth}
  \centering
  \includegraphics[width=0.95\linewidth]{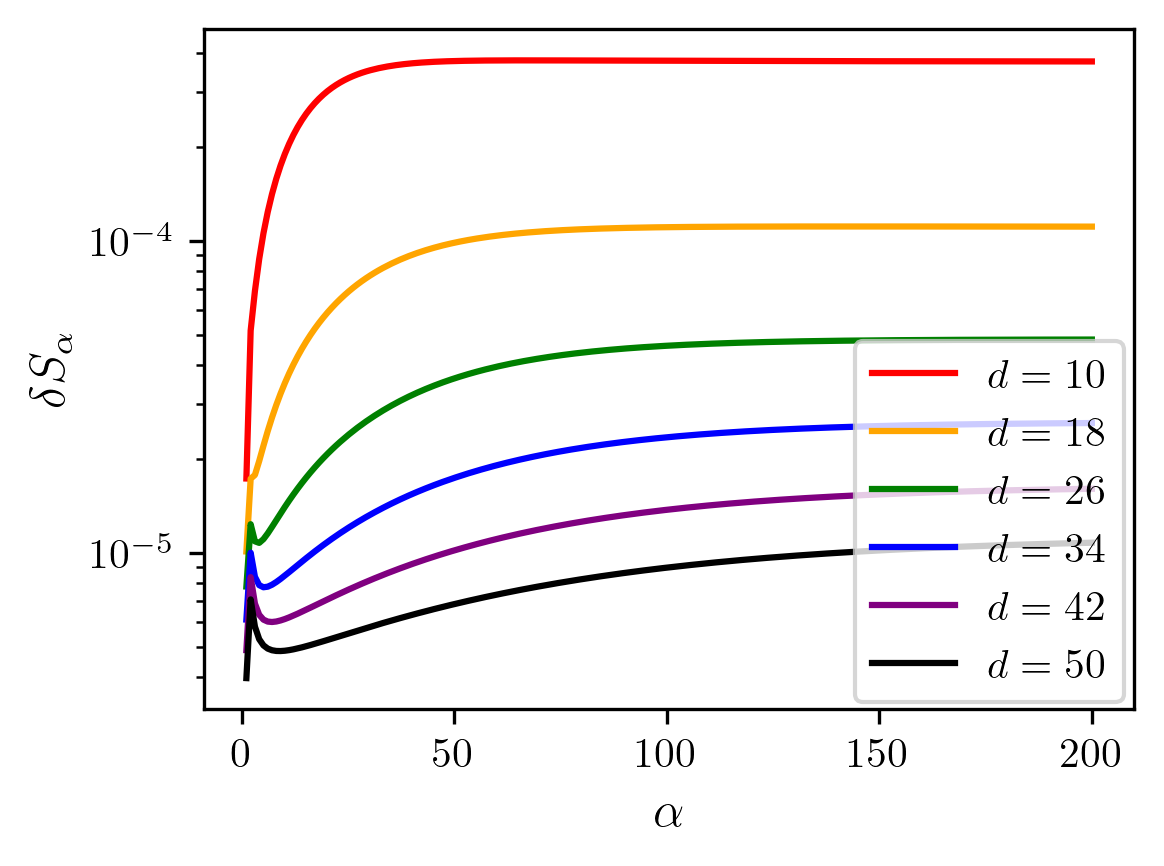}
  \caption{Deviation $\delta S_\alpha$ for different $d$. }
  \label{fig:sfig2}
\end{subfigure}
\caption{R\'eyni entropies $S_\alpha$ of an anti-Krawtchouk chain reduced density matrix $\varrho_A$ and the density matrix $\varrho_T(t_1,t_2)$ based on an affine transformation of the Heun operator with constraints \eqref{eq:c1c2} and with parameters $j = d$ and $L = K = d/2$. Figure~\ref{fig:sfig1} shows the similarities between $S_\alpha(\varrho_A)$ and $S_\alpha(\rho_T(t_1,t_2))$ for all $\alpha$ at $d = 40$. Figure~\ref{fig:sfig2} presents their deviation $\delta S_\alpha$ for different diameters $d$ as a function of the R\'enyi index $\alpha$.}
\label{fig:fig}
\end{figure}

Numerically, we find that the deviation is small relative to $S_\alpha$ for all $\alpha$ when fixing $t_1$ and $t_2$ such that the constraints~\eqref{eq:c1c2} are verified, see Fig.~\ref{fig:fig}. Imposing that the two states have the same average number of particles and von Neumann entropy is thus sufficient to ensure that $\varrho_A$ and $\varrho_T(t_1,t_2)$ share a similar entanglement spectrum.

\section{Conclusion} \label{sec:conclusion}
The scaling behavior of the ground-state entanglement entropy was investigated for a model of free fermions defined on the vertices of a folded cube. In the limit of large diameter, the entanglement entropy was found to obey a strict area law without any logarithmic enhancement. This departure from the behavior observed in free-fermion systems on cubic lattices can be attributed to the intricate geometric structure of folded cubes. Specifically, these structures can be effectively decomposed into a collection of chains, most of which only have a small intersection with the subsystem compared to the graph's diameter. From the perspective of these chains, the thermodynamic limit of folded cubes with $d \rightarrow \infty$ and a fixed aspect ratio $L/d$ maps to a thermodynamic limit of one-dimensional systems having aspect ratios approaching zero, and hence giving no logarithmic enhancement. 

A similar phenomenon and explanation should hold for the thermodynamic limits of free fermions on sequences of other distance-regular graphs with increasing diameter. It would be intriguing to investigate whether the rich symmetries of distance-regular graphs are essential or if similar patterns emerge in a wide range of high-dimensional graphs.

Additionally, we explored the relationship between the entanglement Hamiltonian and the Heun operator. It was observed that the reduced density matrix and a Gaussian state of a Hamiltonian constructed through an affine transformation of the Heun operator have R\'eyni fidelities close to one, provided that both matrices possess equal expectations of particle number and von Neumann entropy. It was also shown that they have similar R\'eyni entropies. It suggests that free-fermion Hamiltonians based on generalized Heun operators offer adequate approximation of entanglement Hamiltonians. Future investigations could look into alternative constraints on the affine transformation so as to possibly find higher fidelities, a more accurate alignment of the R\'enyi entropies and, consequently, a better approximation of the entanglement Hamiltonian.

\subsection*{Acknowledgement}
We thank Riccarda Bonsignori for useful discussion and correspondence. ZM was supported by USRA scholarships from NSERC and FRQNT. PAB holds an Alexander-Graham-Bell scholarship from the Natural Sciences and Engineering Research Council of Canada (NSERC). GP holds a FRQNT and a CRM–ISM postdoctoral fellowship, and acknowledges support from the Mathematical Physics Laboratory of the CRM.  The research of LV is funded in part by a Discovery Grant from NSERC.

\providecommand{\href}[2]{#2}\begingroup\raggedright\endgroup


\providecommand{\href}[2]{#2}\begingroup\raggedright\begin{thebibliography}{10}

\bibitem{amico2008entanglement}
L.~Amico, R.~Fazio, A.~Osterloh, and V.~Vedral, ``Entanglement in many-body
  systems,'' \href{http://dx.doi.org/10.1103/RevModPhys.80.517}{{\em Rev. Mod.
  Phys.} {\bfseries 80}, 517 (2008)}.

\bibitem{laflorencie2016quantum}
N.~Laflorencie, ``Quantum entanglement in condensed matter systems,''
  \href{http://dx.doi.org/10.1016/j.physrep.2016.06.008}{{\em Phys. Rep.}
  {\bfseries 646}, 1 (2016)}.

\bibitem{OAFF02}
A.~Osterloh, L.~Amico, G.~Falci, and R.~Fazio, ``{Scaling of entanglement close
  to a quantum phase transitions},''
  \href{http://dx.doi.org/10.1038/416608a}{{\em Nature} {\bfseries 416}, 608
  (2002)}.

\bibitem{ON02}
T.~J. Osborne and M.~A. Nielsen, ``{Entanglement in a simple quantum phase
  transition},'' \href{http://dx.doi.org/10.1103/PhysRevA.66.032110}{{\em Phys.
  Rev.~A} {\bfseries 66}, 032110 (2002)}.

\bibitem{vidal2003entanglement}
G.~Vidal, J.~I. Latorre, E.~Rico, and A.~Kitaev, ``Entanglement in quantum
  critical phenomena,''
  \href{http://dx.doi.org/10.1103/PhysRevLett.90.227902}{{\em Phys. Rev. Lett.}
  {\bfseries 90}, 227902 (2003)}.

\bibitem{CC04}
P.~Calabrese and J.~L. Cardy, ``{Entanglement entropy and quantum field
  theory},'' \href{http://dx.doi.org/10.1088/1742-5468/2004/06/P06002}{{\em
  J.~Stat. Mech.} P06002 (2004)}.

\bibitem{calabrese2009entanglement}
P.~Calabrese and J.~L. Cardy, ``Entanglement entropy and conformal field
  theory,'' \href{http://dx.doi.org/10.1088/1751-8113/42/50/504005}{{\em J.
  Phys. A: Math. Theor.} {\bfseries 42}, 504005 (2009)}.

\bibitem{kitaev2006topological}
A.~Kitaev and J.~Preskill, ``{Topological entanglement entropy},''
  \href{http://dx.doi.org/10.1103/PhysRevLett.96.110404}{{\em Phys. Rev. Lett.}
  {\bfseries 96}, 110404 (2006)}.

\bibitem{levin2006detecting}
M.~Levin and X.-G. Wen, ``Detecting topological order in a ground state wave
  function,'' \href{http://dx.doi.org/10.1103/PhysRevLett.96.110405}{{\em Phys.
  Rev. Lett.} {\bfseries 96}, 110405 (2006)}.

\bibitem{cc-05}
P.~Calabrese and J.~L. Cardy, ``{Evolution of entanglement entropy in
  one-dimensional systems},''
  \href{http://dx.doi.org/10.1088/1742-5468/2005/04/p04010}{{\em
  J.~Stat.~Mech.} P04010 (2005)}.

\bibitem{fc-08}
M.~Fagotti and P.~Calabrese, ``{Evolution of entanglement entropy following a
  quantum quench: Analytic results for the XY chain in a transverse magnetic
  field},'' \href{http://dx.doi.org/10.1103/PhysRevA.78.010306}{{\em Phys. Rev.
  A} {\bfseries 78}, 010306 (2008)}.

\bibitem{GE15}
C.~Gogolin and J.~Eisert, ``{Equilibration, thermalisation, and the emergence
  of statistical mechanics in closed quantum systems},''
  \href{http://dx.doi.org/10.1088/0034-4885/79/5/056001}{{\em Rep. Prog. Phys.}
  {\bfseries 79}, 056001 (2016)}.

\bibitem{ac-17}
V.~Alba and P.~Calabrese, ``Entanglement and thermodynamics after a quantum
  quench in integrable systems,''
  \href{http://dx.doi.org/10.1073/pnas.1703516114}{{\em Proceedings of the
  National Academy of Sciences} {\bfseries 114}, 7947 (2017)}.

\bibitem{gioev2006entanglement}
D.~Gioev and I.~Klich, ``{Entanglement entropy of fermions in any dimension and
  the Widom conjecture},''
  \href{http://dx.doi.org/https://doi.org/10.1103/PhysRevLett.96.100503}{{\em
  Phys. Rev. Lett.} {\bfseries 96}, 100503 (2006)}.

\bibitem{li2006scaling}
W.~Li, L.~Ding, R.~Yu, T.~Roscilde, and S.~Haas, ``Scaling behavior of
  entanglement in two- and three-dimensional free-fermion systems,''
  \href{http://dx.doi.org/https://doi.org/10.1103/PhysRevB.74.073103}{{\em
  Phys. Rev. B} {\bfseries 74}, 073103 (2006)}.

\bibitem{bernard2023entanglement}
P.-A. Bernard, N.~Cramp{\'e}, and L.~Vinet, ``{Entanglement of free fermions on
  {H}amming graphs},''
  \href{http://dx.doi.org/https://doi.org/10.1016/j.nuclphysb.2022.116061}{{\em
  Nucl. Phys. B} {\bfseries 986}, 116061 (2023)}.

\bibitem{parez2022multipartite}
G.~Parez, P.-A. Bernard, N.~Cramp{\'e}, and L.~Vinet, ``{Multipartite
  information of free fermions on Hamming graphs},''
  \href{http://dx.doi.org/https://doi.org/10.1016/j.nuclphysb.2023.116157}{{\em
  Nucl. Phys. B} {\bfseries 990}, 116157 (2023)}.

\bibitem{bernard2021entanglement}
P.-A. Bernard, N.~Crampé, and L.~Vinet, ``{{Entanglement of free fermions on
  Johnson graphs}},'' \href{http://dx.doi.org/10.1063/5.0099879}{{\em J. Math.
  Phys.} {\bfseries 64}, 061903 (2023)}.

\bibitem{brown2013hypercubes}
G.~M. Brown, ``Hypercubes, {L}eonard triples and the anticommutator spin
  algebra,'' \href{http://arxiv.org/abs/1301.0652}{{\ttfamily arXiv:1301.0652
  [math.CO]}}.

\bibitem{t1}
P.~Terwilliger, ``{The subconstituent algebra of an association scheme ({P}art
  {I})},''
  \href{http://dx.doi.org/https://doi.org/10.1023/A:1022494701663}{{\em J.
  Algebr. Comb.} {\bfseries 1}, 363 (1992)}.

\bibitem{t2}
P.~Terwilliger, ``The subconstituent algebra of an association scheme ({P}art
  {I}{I}),''
  \href{http://dx.doi.org/https://doi.org/10.1023/A:1022480715311}{{\em J.
  Algebr. Comb.} {\bfseries 2}, 73 (1993)}.

\bibitem{t3}
P.~Terwilliger, ``The subconstituent algebra of an association scheme ({P}art
  {I}{I}{I}),''
  \href{http://dx.doi.org/https://doi.org/10.1023/A:1022415825656}{{\em J.
  Algebr. Comb.} {\bfseries 2}, 177 (1993)}.

\bibitem{eisler2013free}
V.~Eisler and I.~Peschel, ``Free-fermion entanglement and spheroidal
  functions,'' \href{http://dx.doi.org/10.1088/1742-5468/2013/04/P04028}{{\em
  J.~Stat.~Mech} P04028 (2013)}.

\bibitem{eisler2017analytical}
V.~Eisler and I.~Peschel, ``{Analytical results for the entanglement
  {H}amiltonian of a free-fermion chain},''
  \href{http://dx.doi.org/10.1088/1751-8121/aa76b5}{{\em J. Phys. A: Math.
  Theor.} {\bfseries 50}, 284003 (2017)}.

\bibitem{eisler2018properties}
V.~Eisler and I.~Peschel, ``Properties of the entanglement {H}amiltonian for
  finite free-fermion chains,''
  \href{http://dx.doi.org/10.1088/1742-5468/aace2b}{{\em J.~Stat.~Mech.} 104001
  (2018)}.

\bibitem{parez2022symmetry}
G.~Parez, ``Symmetry-resolved {R}{\'e}nyi fidelities and quantum phase
  transitions,''
  \href{http://dx.doi.org/https://doi.org/10.1103/PhysRevB.106.235101}{{\em
  Phys. Rev. B} {\bfseries 106}, 235101 (2022)}.

\bibitem{brouwer2012distance}
A.~E. Brouwer, W.~H. Haemers, A.~E. Brouwer, and W.~H. Haemers,
  \href{http://dx.doi.org/https://doi.org/10.1007/978-3-642-74341-2}{{\em
  Distance-regular graphs}}.
\newblock Springer, 2012.

\bibitem{peschel2003calculation}
I.~Peschel, ``Calculation of reduced density matrices from correlation
  functions,'' \href{http://dx.doi.org/10.1088/0305-4470/36/14/101}{{\em J.
  Phys. A: Math. Gen.} {\bfseries 36}, L205 (2003)}.

\bibitem{genest}
V.~X. Genest, L.~Vinet, G.-F. Yu, and A.~Zhedanov, {\em Supersymmetry of the
  {Q}uantum {R}otor, in {F}rontiers in {O}rthogonal {P}olynomials and
  $q$-{S}eries}, \href{http://dx.doi.org/10.1142/9789813228887_0015}{ch.~15,
  pp.~291--305}.
\newblock World Scientific, (2018).

\bibitem{Crampe:2019upj}
N.~Cramp\'e, R.~I. Nepomechie, and L.~Vinet, ``{Free-Fermion entanglement and
  orthogonal polynomials},''
  \href{http://dx.doi.org/10.1088/1742-5468/ab3787}{{\em J.~Stat.~Mech.} 093101
  (2019)}.

\bibitem{FA20}
F.~Finkel and A.~Gonz{\'a}lez-L{\'o}pez, ``{Inhomogeneous XX spin chains and
  quasi-exactly solvable models},''
  \href{http://dx.doi.org/10.1088/1742-5468/abb237}{{\em J.~Stat.~Mech.} 093105
  (2020)}.

\bibitem{FA21}
F.~Finkel and A.~Gonz{\'a}lez-L{\'o}pez, ``{Entanglement entropy of
  inhomogeneous XX spin chains with algebraic interactions},''
  \href{http://dx.doi.org/10.1007/JHEP12(2021)184}{{\em JHEP} 1 (2021)}.

\bibitem{Crampe:2021}
N.~Cramp\'e, R.~I. Nepomechie, and L.~Vinet, ``{Entanglement in fermionic
  chains and bispectrality},''
  \href{http://dx.doi.org/10.1142/S0129055X21400018}{{\em Rev. Math. Phys.}
  {\bfseries 33}, 2140001 (2021)}.

\bibitem{bernard2022entanglement}
P.-A. Bernard, N.~Cramp{\'e}, R.~I. Nepomechie, G.~Parez, L.~P. d'Andecy, and
  L.~Vinet, ``Entanglement of inhomogeneous free fermions on hyperplane
  lattices,''
  \href{http://dx.doi.org/https://doi.org/10.1016/j.nuclphysb.2022.115975}{{\em
  Nucl. Phys. B} {\bfseries 984}, 115975 (2022)}.

\bibitem{bernard2022computation}
P.-A. Bernard, G.~Carcone, N.~Crampe, and L.~Vinet, ``{Computation of
  entanglement entropy in inhomogeneous free fermions chains by algebraic Bethe
  ansatz},'' \href{http://arxiv.org/abs/2212.09805}{{\ttfamily arXiv:2212.09805
  [math-ph]}}.

\bibitem{DSVC17}
J.~Dubail, J.-M. St\'ephan, J.~Viti, and P.~Calabrese, ``{Conformal field
  theory for inhomogeneous one-dimensional quantum systems: the example of
  non-interacting Fermi gases},''
  \href{http://dx.doi.org/10.21468/SciPostPhys.2.1.002}{{\em SciPost Phys.}
  {\bfseries 2}, 002 (2017)}.

\bibitem{grunbaum2018algebraic}
F.~A. Gr{\"u}nbaum, L.~Vinet, and A.~Zhedanov, ``Algebraic {H}eun operator and
  band-time limiting,''
  \href{http://dx.doi.org/https://doi.org/10.1007/s00220-018-3190-0}{{\em
  Commun. Math. Phys.} {\bfseries 364}, 1041 (2018)}.

\bibitem{eisler2019continuum}
V.~Eisler, E.~Tonni, and I.~Peschel, ``{On the continuum limit of the
  entanglement Hamiltonian},''
  \href{http://dx.doi.org/10.1088/1742-5468/ab1f0e}{{\em J. Stat. Mech} 073101
  (2019)}.

\bibitem{bernard2021heun}
P.-A. Bernard, N.~Cramp{\'e}, D.~Shaaban~Kabakibo, and L.~Vinet, ``Heun
  operator of lie type and the modified algebraic {B}ethe ansatz,''
  \href{http://dx.doi.org/https://doi.org/10.1063/5.0041097}{{\em J. Math.
  Phys.} {\bfseries 62}, 083501 (2021)}.

\bibitem{BE23pc}
R.~Bonsignori and V.~Eisler, ``{Private communication},''   (2023).

\bibitem{U76}
A.~Uhlmann, ``{The `transition probability' in the state space of a
  $\ast$-algebra},'' \href{http://dx.doi.org/10.1016/0034-4877(76)90060-4}{{\em
  Rep. Math. Phys.} {\bfseries 9}, 273 (1976)}.

\bibitem{J94}
R.~Jozsa, ``{Fidelity for Mixed Quantum States},''
  \href{http://dx.doi.org/10.1080/09500349414552171}{{\em J.~Mod.~Opt.}
  {\bfseries 41}, 2315 (1994)}.

\bibitem{QSLZS06}
H.~T. Quan, Z.~Song, X.~F. Liu, P.~Zanardi, and C.~P. Sun, ``{Decay of
  Loschmidt Echo Enhanced by Quantum Criticality},''
  \href{http://dx.doi.org/10.1103/PhysRevLett.96.140604}{{\em Phys.~Rev.~Lett.}
  {\bfseries 96}, 140604 (2006)}.

\bibitem{ZB08}
H.-Q. Zhou and J.~P. Barjaktarevi{\v c}, ``{Fidelity and quantum phase
  transitions},'' \href{http://dx.doi.org/10.1088/1751-8113/41/41/412001}{{\em
  J.~Phys.~A: Math. Theor.} {\bfseries 41}, 412001 (2008)}.

\bibitem{G10}
S.-J. Gu, ``{Fidelity approach to quantum phase transitions},''
  \href{http://dx.doi.org/10.1142/S0217979210056335}{{\em Int.~J.~Mod.~Phys.~B}
  {\bfseries 24}, 4371 (2010)}.

\bibitem{DS11}
J.~Dubail and J.-M. St\'ephan, ``{Universal behavior of a bipartite fidelity at
  quantum criticality},''
  \href{http://dx.doi.org/10.1088/1742-5468/2011/03/L03002}{{\em
  J.~Stat.~Mech.} L03002 (2011)}.

\bibitem{SD13}
J.-M. St\'ephan and J.~Dubail, ``{Logarithmic corrections to the free energy
  from sharp corners with angle $2\pi$},''
  \href{http://dx.doi.org/10.1088/1742-5468/2013/09/P09002}{{\em J.~Stat.
  Mech.} P09002 (2013)}.

\bibitem{hagendorf2017open}
C.~Hagendorf and J.~Li{\'e}nardy, ``Open spin chains with dynamic lattice
  supersymmetry,'' \href{http://dx.doi.org/10.1088/1751-8121/aa67ff}{{\em J.
  Phys. A: Math. Theor.} {\bfseries 50}, 185202 (2017)}.

\bibitem{PMDR19}
G.~Parez, A.~Morin-Duchesne, and P.~Ruelle, ``Bipartite fidelity of critical
  dense polymers,'' \href{http://dx.doi.org/10.1088/1742-5468/ab310f}{{\em
  J.~Stat.~Mech.} 103101 (2019)}.

\bibitem{MDPL21}
A.~Morin-Duchesne, G.~Parez, and J.~Li\'enardy, ``Bipartite fidelity for models
  with periodic boundary conditions,''
  \href{http://dx.doi.org/10.1088/1742-5468/abc1eb}{{\em J.~Stat.~Mech.} 023101
  (2021)}.

\bibitem{HP21}
C.~Hagendorf and G.~Parez, ``{On the logarithmic bipartite fidelity of the open
  {X}{X}{Z} spin chain at $\Delta =-1/2$},''
  \href{http://dx.doi.org/10.21468/SciPostPhys.12.6.199}{{\em SciPost Phys.}
  {\bfseries 12}, 199 (2022)}.

\bibitem{chuang}
M.~A. Nielsen and I.~L. Chuang,
  \href{http://dx.doi.org/10.1017/CBO9780511976667}{{\em Quantum Computation
  and Quantum Information: 10th Anniversary Edition}}.
\newblock Cambridge University Press, 2010.

\bibitem{carrasco2017duality}
J.~A. Carrasco, F.~Finkel, A.~Gonzalez-Lopez, and P.~Tempesta, ``A duality
  principle for the multi-block entanglement entropy of free fermion systems,''
  \href{http://dx.doi.org/https://doi.org/10.1038/s41598-017-09550-1}{{\em Sci.
  Rep.} {\bfseries 7}, 11206 (2017)}.

\end{thebibliography}\endgroup
\newpage
\appendix

\section{Anti-Krawtchouk polynomials}
\label{appA}

The monic anti-Krawtchouk polynomials satisfy the following three-term recurrence relation,
\begin{equation}
    xP_n(x)=P_{n+1}(x)-(A_n+B_n)P_n(x)+A_{n-1}C_nP_{n-1}(x),
\end{equation}
where
\begin{equation}
    A_n=\frac{(-1)^{n+N+1}(N+1)+n+1}{4}, \qquad C_n=\begin{cases}
    0, & \quad n=0 \\ \frac{(-1)^{N+n}(N+1)-n}{4}, & \quad n\neq 0. 
    \end{cases}
\end{equation}

The polynomials are given in terms of $_4F_3$ generalized hypergeometric series as
\begin{equation}
    P_n(x)=\mathcal{A}_{n} \times \begin{cases}
    \, _4F_3\lrp{\substack{ \frac{-(n-1)}{2},\ \frac{n+1}{2},\ \frac{x}{2}+\frac{1}{4},\ -\frac{x}{2}+\frac{3}{4} \\ 1-(-1)^N\lrp{\frac{N+1}{2}},\ \frac{1}{2},\ \frac{1}{2}+(-1)^N\lrp{\frac{N+1}{2}}};1}\\[.4cm] \quad
        -\frac{(n+1)(\frac{x}{2}+\frac{1}{4})}{\frac{1}{2}+(-1)^N\lrp{\frac{N+1}{2}}}\, _4F_3\lrp{ \substack{ -\frac{(n-1)}{2},\ \frac{n}{2}+1,\ \frac{x}{2}+\frac{5}{4},\ -\frac{x}{2}+\frac{3}{4} \\ 1-(-1)^N\lrp{\frac{N+1}{2}},\ \frac{3}{2},\ \frac{3}{2}+(-1)^N\lrp{\frac{N+1}{2}}};1 }, & \quad n\tx{ odd,} \\ \\[.5cm]
        \, _4F_3\lrp{\substack{ \frac{-n}{2},\ \frac{n}{2}+1,\ \frac{x}{2}+\frac{1}{4},\ -\frac{x}{2}+\frac{3}{4} \\ 1-(-1)^N\lrp{\frac{N+1}{2}},\ \frac{1}{2},\ \frac{1}{2}+(-1)^N\lrp{\frac{N+1}{2}}};1}\\[.4cm] \quad
        +\frac{n(\frac{x}{2}+\frac{1}{4})}{\frac{1}{2}+(-1)^N\lrp{\frac{N+1}{2}}} 
        \, _4F_3\lrp{\substack{1-\frac{n}{2},\ 1+\frac{n}{2},\ \frac{x}{2}+\frac{5}{4},\ -\frac{x}{2}+\frac{3}{4} \\ 1-(-1)^N\lrp{\frac{N+1}{2}},\ \frac{3}{2},\ \frac{3}{2}+(-1)^N\lrp{\frac{N+1}{2}}};1}, & \quad n\tx{ even,}
    \end{cases}
\end{equation}
where $\mathcal{A}_n$ is a coefficient given by
\begin{equation}
    \mathcal{A}_{n}=\begin{cases}
    \frac{\alpha_N+1/2}{-(n+1)}{\lrp{1-\alpha_N}_{\frac{n-1}{2}}\lrp{\frac{3}{2}}_{\frac{n-1}{2}}\lrp{\frac{3}{2}+\alpha_N}_{\frac{n-1}{2}}}{\lrp{\frac{n+3}{2}}_{\frac{n-1}{2}}^{-1}}, & \quad n\tx{ odd,} \\[.4cm]
   {\lrp{1-\alpha_N}_{\frac{n}{2}}\lrp{1/2}_{\frac{n}{2}}\lrp{\frac{n}{2}+\alpha_N}_{\frac{n}{2}}}{\lrp{\frac{n}{2}+1}_{\frac{n}{2}}^{-1}}, & \quad n\tx{ even,}
    \end{cases}
\end{equation}
and $(a)_k \equiv (a) (a +1) \dots (a + k -1)$ is the Pochhammer symbol. 

The monic anti-Krawtchouk polynomials are orthogonal on the grid
\begin{equation}
    x_k=(-1)^k\lrp{k+\frac{1}{2}},
\end{equation}
with the weight function
\begin{equation}
    \Omega_k=\begin{cases}
    \lrp{\frac{\lrp{-N/2}_{k/2}\lrp{N/2+3/2}_{k/2}}{\lrp{1+N/2}_{k/2}\lrp{1/2-N/2}_{k/2}}}^{(-1)^N} , & \quad n\tx{ even,} \\[.4cm]
    \lrp{\frac{\lrp{-N/2}_{(k+1)/2}\lrp{N/2+3/2}_{(k-1)/2}}{\lrp{1+N/2}_{(k+1)/2}\lrp{1/2-N/2}_{(k-1)/2}}}^{(-1)^N},  & \quad n\tx{ odd.}
    \end{cases}
\end{equation}

Indeed, their non-monic counterpart $\hat{P}_n(x)$ defined by
\begin{equation}
    \hat{P}_n(x)=\frac{4^n}{\sqrt{(N+2)_n(N+1-n)_n}} P_n(x),
\end{equation}
verify the following two orthogonality relations,
\begin{equation}
    \sum_{k=0}^N \Omega_k\hat{P}_n(x_k)\hat{P}_m(x_k)=\Phi_N\delta_{nm}, \qquad  \sum_{n=0}^N \Omega_k\hat{P}_n(x_k)\hat{P}_n(x_{\ell})=\Phi_N\delta_{k\ell},
\end{equation}
where $\Phi_N$ is a normalisation factor given by
\begin{equation}
    \Phi_N=\begin{cases}
    _3F_2\lrp{\substack{-N/2,\ N/2+3/2,\ 1\\ 1+N/2,\ 1/2-N/2\ };1}  +\lrp{\frac{N}{N+2}}\, _3F_2\lrp{\substack{-N/2+1,\ N/2+3/2,\ 1\\ 2+N/2,\ 1/2-N/2};1}, & \quad N\tx{ even,} \\[.5cm]
    _3F_2\lrp{\substack{1+N/2,\ 1/2-N/2,\ 1\\ -N/2,\ N/2+3/2};1} +\frac{2+N}{N}\, _3F_2\lrp{\substack{2+N/2,\ -N/2+1/2,\ 1\\ -N/2+1,\ N/2+3/2};1} ,& \quad N\tx{ odd.}
    \end{cases}
\end{equation}
They further solve the recurrence relation \eqref{rr1}, 
\begin{equation}
    y\hat{P}_n(y)=U_{n+1}\hat{P}_{n+1}(y)+B_n\hat{P}_n(y)+U_n\hat{P}_{n-1}(y)
\end{equation}
where
\begin{equation}
    U_n=\sqrt{\frac{(N+1+n)(N+1-n)}{4}}, \qquad B_n=\begin{cases}
    (-1)^N\lrp{\frac{N+1}{2}}, &\quad n=0, \\
    0, &\quad n\neq 0.
    \end{cases}
\end{equation}

Since the bases $\{\ket{j,r,n}_3 : 0 \leqslant n \leqslant j\}$ and $\{\ket{j,r,k}_1: 0 \leqslant n \leqslant j\}$ of $\mathcal{V}_{j,r}$ are respectively orthonormal, the overlaps $Q_{k,n}$ are given by
\begin{equation}
    Q_{k,n} = \sqrt{\frac{\Omega_k}{\Phi_N}}\hat{P}_n(x_k).
\end{equation}
\end{document}